\documentstyle{mn}
\input{psfig.sty}
\psfull
\newcommand{\etal}{{et al.~}}

\newcommand{\beq}{\begin{equation}}
\newcommand{\eeq}{\end{equation}}

\def\gtsima{$\; \buildrel > \over \sim \;$}
\def\ltsima{$\; \buildrel < \over \sim \;$}
\def\prosima{$\; \buildrel \propto \over \sim \;$}
\def\gsim{\lower.7ex\hbox{\gtsima}}
\def\lsim{\lower.7ex\hbox{\ltsima}}
\def\simgt{\lower.7ex\hbox{\gtsima}}
\def\simlt{\lower.7ex\hbox{\ltsima}}
\def\simpr{\lower.7ex\hbox{\prosima}}

\newcommand{\apj}{ApJ}

\newcommand{\mnras}{MNRAS}

\newdimen\hssize
\newdimen\hdsize
\newcommand{\revise}{\textrm}
\begin{document}

%%%%%%%%%%%%%%%%%%%%%%%%%%%%%%%%%%%%%%%%%%%%%%%%%%%%%%%%%%%%%%%%%%%%%%%%%%

\title[Equilibrium Initialization and Stability of Three-Dimensional Gas Disks] 
      {Equilibrium Initialization and Stability of Three-Dimensional Gas Disks}
\author[Wang et al.]
       {\parbox[t]{\textwidth}{
        Hsiang-Hsu Wang$^{1,2}$\thanks{E-mail: hhwang@mpia.de}, 
        Ralf S. Klessen$^{2,5}$,
        Cornelis P. Dullemond$^{1}$,
        Frank C. van den Bosch$^{4}$,
        Burkhard Fuchs$^{3}$
        }\\        
        \vspace*{3pt} \\
       $^1$Max-Planck-Institut f\"ur Astronomie, K\"onigstuhl 17, D-69117
           Heidelberg, Germany\\
       $^2$Zentrum f\"ur Astronomie der Universit\"at Heidelberg, Institut f\"ur Theoretische Astrophysik, Albert-Ueberle-Strasse 2, 69120 Heidelberg, Germany\\
       $^3$Zentrum f\"ur Astronomie der Universit\"at Heidelberg, Astronomisches Rechen-Institut, M\"onchhofstr. 12-14, 69120 Heidelberg, Germany\\
       $^4$Department of Physics and Astronomy, University of Utah, 115 South 1400 East, Salt Lake City, UT 84112, U.S.A. \\
       $^5$Kavli Institute for Particle Astrophysics and Cosmology, Stanford University, Menlo Park, CA 94025, U.S.A.\\
       \\
       Accepted by MNRAS 2010 April 28
       }
%%%%%%%%%%%%%%%%%%%%%%%%%%%%%%%%%%%%%%%%%%%%%%%%%%%%%%%%%%%%%%%%%%%%%%%%%%

\date{}
\pagerange{\pageref{firstpage}--\pageref{lastpage}}
\pubyear{2009}

\maketitle

\label{firstpage}

%%%%%%%%%%%%%%%%%%%%%%%%%%%%%%%%%%%%%%%%%%%%%%%%%%%%%%%%%%%%%%%%%%%%%%%%%%

\begin{abstract}
  We present a new systematic way of setting up galactic gas disks
  based on the assumption of detailed hydrodynamic equilibrium. To do
  this, we need to specify the density distribution and the velocity
  field which supports the disk. We first show that the required
  circular velocity has no dependence on the height above or below the
  midplane so long as the gas pressure is a function of density
  only. The assumption of disks being very thin enables us to decouple
  the vertical structure from the radial direction. Based on that, the
  equation of hydrostatic equilibrium together with the reduced
  Poisson equation leads to two sets of second-order non-linear
  differential equation, which are easily integrated to set-up a
  stable disk. We call one approach `density method' and the other one
  `potential method'. Gas disks in detailed balance are especially
  suitable for investigating the onset of the gravitational
  instability. We revisit the question of global, axisymmetric
  instability using fully three-dimensional disk simulations. The
  impact of disk thickness on the disk instability and the formation
  of spontaneously induced spirals is studied systematically with or
  without the presence of the stellar potential. In our models, the
  numerical results show that the threshold value for disk instability
  is shifted from unity to 0.69 for self-gravitating thick disks and
  to 0.75 for combined stellar and gas thick disks. The simulations
  also show that self-induced spirals occur in the correct regions and
  with the right numbers as predicted by the analytic theory.
\end{abstract}

%%%%%%%%%%%%%%%%%%%%%%%%%%%%%%%%%%%%%%%%%%%%%%%%%%%%%%%%%%%%%%%%%%%%%%%%%%


%%%%%%%%%%%%%%%%%%%%%%%%%%%%%%%%%%%%%%%%%%%%%%%%%%%%%%%%%%%%%%%%%%%%%%%%%%

\section{Introduction}

The stability of gas disks plays an important role in governing the
structure of disk galaxies and in regulating their star formation
rate. Although important insights can be obtained using perturbation
theory (Toomre 1964, Lin \& Shu 1964, Rafikov 2001), the onset of
stability and its impact on the star formation and evolution of gas
disks is best studied using hydrodynamical simulations. These can
follow the non-linear behavior of the system, which cannot be
addressed by linear analysis. With the recent advances in computing
power and the development of new numerical techniques, we are now in a
good position to treat a three-dimensional, isolated galaxy
self-consistently.

However, in order for a stability analysis to be meaningful and
reliable, it is of paramount importance that one can specify
equilibrium initial conditions. After all, if the initial disk is not
in equilibrium, its relaxation during the first time-steps of the
simulation may trigger instabilities that are of little relevance for
our understanding of the stability of disk galaxies.  Unfortunately,
no analytical solution is known for the density, velocity field and
temperature of a three-dimensional gas disk in hydrostatic equilibrium
in the external potential of a dark matter halo and/or a stellar
disk. Consequently, previous hydrodynamical simulations have either
started from non-equilibrium initial conditions, or have resorted to
iterative techniques to set-up the initial conditions, at the cost of
having little control over the resulting equilibrium configuration. 
In this paper we present a new method that allows one to compute
the density and velocity structure of a realistic, isothermal,
three-dimensional gas disk in hydrostatic equilibrium in an abritrary
external potential.
   
Hydrostatic equilibrium implies a balance between gravity and
pressure. Gravity includes the self-gravity of the disk plus that of
external components (i.e. dark matter halo, bulge, stellar disk, etc),
while the pressure is given by an equation of state $p=p(\rho_{\rm
g},T)$, with $p$ being the gas pressure, $\rho_{\rm g}$ the gas
density and $T$ the temperature. The challenge is to find a $\rho_{\rm
g}$, $T$ and the velocity field, $\vec{v}$, such that the system is
self-consistent (i.e., obeys the Poisson equation) and in hydrostatic
equilibrium.
   
In the case of an isothermal, axisymmetric, perfectly self-gravitating
disk (i.e., no external potential), the equilibrium disk has a ${\rm
sech}^2$ distribution (Spitzer 1942) in the vertical direction, with a
scale-height that is proportional to $\sqrt{c^2_{\rm s} / \rho_{\rm
g}(R,z=0)}$, where $c_{\rm s}$ is the sound speed.  Here, cylindrical
coordinates, $(R,\phi, z)$, are used to describe the density field.
This immediately shows that since $\rho_{\rm g}(R,z=0)$ is typically a
decreasing function of radius, one generally expects the scale-height
to be a function of $R$. In particular, in the case of a globally
isothermal disk, the sound speed $c^2_{\rm s} \propto T$ is constant
in space, giving rise to a flaring disk, i.e., the scale-height
increases with increasing $R$ (Narayan \& Jog 2002, hereafter NJ02;
Agertz \etal 2009).  Alternatively, if we want to initialize a disk
with a constant scale-height, a radial temperature gradient needs to
be introduced. Tasker \& Bryan (2006) initialize their disks to be
isothermal and to have a constant scale-height. As indicated above,
this cannot be an equilibrium configuration. Consequently, the disk is
expected to experience an unavoidable relaxation process which makes
the initialization not well-controlled and might potentially
contaminate the physics, e.g., star formation, gas dynamics etc., of
interest. Agertz \etal (2009) set-up their isothermal disks based on
the local total surface density of gas plus dark matter. Although the
scale-height of their initial disk changes with radius, the local total
surface density is not defined in a mathematical way and therefore
elusive. In addition, their surface density does not follow an
exponential profile. 
    
\revise{An important assumption underlying Spitzer's analysis is that
the radial variation in the potential is negligible compared to that
in the vertical direction. This assumption is supported by observation
that disks typically have vertical scale-heights that are an order of
magnitude smaller than their radial scale-length (van der Kruit \& Searle 1981a,b). A well studied
example is the Milky Way, whose scale-height is less than 200 pc for
the cold gas (Jackson \& Kellman 1974; Lockman 1984; Sanders \etal 1984; Wouterloot \etal 1990; 
see also Narayan \& Jog 2002) and roughly 300 pc for the stars in the 
Solar neighborhood (Binney \& Tremaine 2008, Kent, Dame \& Fazio 1991), 
compared to a radial scale-length of $\sim 3.5$ kpc. Throughout this paper we
therefore follow Spitzer and consider disks to be `thin', allowing us
to treat their radial and vertical structure separately. Hence, we
caution that our method is not valid for thick disk structures.
However, since we are mainly concerned with cold gas disks in this
paper, this restriction is of little importance.}
    
Springel, Matteo \& Hernquist (2005) introduce a flexible solution for
initializing a gas disk self-consistently. Basically, they solve
Eq. (2), Eq. (3) and Eq. (24) (see Section 2) iteratively. First, they
deploy a number of particles (say, $2048 \times 64 \times 64$) on a
distorted grid structure in the radial, the azimuthal and the vertical
directions. Unlike the live particles, these particles are simply used
as markers for mass distribution. Second, they compute the joint
total potential and the resulting force field numerically with a
hierarchical multipole expansion based on a tree code. Third, given
the potential just computed, integrating Eq. (2) for a given midplane
volume density, $\rho_{\rm g}(R,z=0)$, gives the vertical structure of
density.  Fourth, adjust the midplane volume density to fulfill
Eq. (24). Repeat the procedure between the second step and the fourth
until the result converges.

Although this description is quite general and flexible, for several
reasons, this is not commonly used in the grid-based codes which are
featured with adaptive-mesh refinement (AMR). The first and also the most
fundamental one is that the grid structure is normally unknown before
we actually initialize the disk. Except the uniform-grid
initialization, the grid structure is automatically generated based on
the criterion for refinement.  Second, for a fully parallelized
code, the initial data is distributed over different processors and
memory storages. This means that the data exchange between processors
is necessary in order to fully compute the joint total potential. The
situation becomes even more technically challenging when initializing
with AMR. Third, The vertical structure of the gas disk depends only
on the vertical potential difference (see Eq. (7) and Eq. (9)
below). A description of the equatorial potential is enough for
specifying the velocity field (See Eq. (13), Eq. (29) and the results
shown in Sec. 3). In general, given the density distribution computed
by the methods introduced in Section 2.2 together with the conclusion
in Section 2.1, we are allowed to acquire the exact velocity field by
Eq. (A.17) in Casertano (1983). Fully solving the Poisson equation becomes
not necessary. Fourth, initializing a disk over distributed memories
allow us to deal with a larger data set which cannot be fully
contained in a single memory storage.

We propose a simple but very effective way of initializing a
three-dimensional gas disk. This method can be easily incorporated
into any existing code based on either a Lagrangian or Eulerian
approach. No data exchange between processors is needed. Vertical
density profile is obtained self-consistently without solving the full
Possion equation. We implement these ideas with the adaptive mesh refinement
magnetohydrodynamics code RAMSES (Teyssier 2002)
and apply our concepts to probe the onset of the disk instability. We
modify the dispersion relation for the infinitesimally thin disk (Lin
\& Shu 1964) to be able to treat thick disks (Goldreich \& Lynden-Bell
1965; Kim \& Ostriker 2002a, 2006; Shetty \& Ostriker 2006; Lisker \&
Fuchs 2009). The threshold value $Q_{\rm th}$ is then obtained
semi-analytically. Previous studies on this subject are either focused
on a small patch of a galaxy (2D/3D: Kim \& Ostriker 2002a) or are
globally two-dimensional but with the reduction of gravity included in
the governing equations (Shetty \& Ostriker 2006). In this paper, we
revisit the subject as a test of our fully three-dimensional isolated
galaxy models. Models with or without stellar potential are
investigated.

Galactic disks are comprised of stars and gas. Both components are
coupled to each other via the Poisson equation. Since the stellar disk
dominates the mass budget within the luminous disk, its presence has
great impact on the scale-height of the gas disk as described in
NJ02. A balanced initial condition depends not only on the correct
vertical structure but also on the correct rotation velocity. To
specify the rotation velocity needed, the mass enclosed within a
certain radius must be under control. Although it is common practice
to specify the functional form of the volume densities of 3D disks, we
show that because of the flaring disk this typically results in a
surface density profile that contains a central `hole' (Agertz \etal
2009). This problem can be trivially avoided by specifying the desired
surface density profile instead. We show that the corresponding volume
density can easily be obtained using a simple iterative scheme. The
surface density of the total gas (${\rm HI}+{\rm H}_2$) from
observation (Leroy \etal 2008) typically follows an expotential
profile in disk galaxies. This profile gives an analytic description
of the total mass enclosed within a radius as well as a reasonable
approximation for velocity field as shown by Eq. (29) below (Binney \&
Tremaine 2008).

Describing the stellar disk with a fixed background potential is at
best an approximation to reality. The interaction between live stellar
disk and gas can potentially destabilize the system (Rafikov 2001; 
Li, Mac Low \& Klessen 2005a, 2005b, 2006; Kim \& Ostriker 2007). After all, 
the gas is cold compared to the stellar disk and has highly non-linear 
response to the asymmetric stellar potential. The gravitational interplay 
between the collisionless stars and dissipative gas is important for 
a number of key questions in galactic dynamics. For example, what is the 
physical origin of grand design spirals? Or what initiates and regulates 
the formation of stars? Having access to well-controlled initial and 
environmental conditions is a prerequisite to discovering their causes.

This paper is organized as follows. The ideas of initializing a gas
disk are outlined in Section 2. Details of the simulation parameters
and test runs are described in Section 3. Axisymmetric instability of
the disk is revisited in Section 4. The self-induced spirals due to
swing amplification will be discussed in Section 5. A brief summary
and the possible extension of this work is put in Section 6.

\section{Formulation of Equations}

In this Section, we develop the required relations and equations to
immerse a 3D gas disk in a preexisting static potential. Assuming that
the gas disk and the preexisting potential share the same symmetry
axis, cylindrical coordinates, $(R,\phi,z)$, are adopted to formulate
the dynamics of the system. Axial-symmetry enables us to discard the
terms describing the variation in azimuthal direction, i.e.,
$\partial/\partial \phi=0$. A gas disk which is in detailed balance
should fulfill the following equations:
\begin{eqnarray}
\frac{1}{\rho_{\rm g}}\frac{\partial p}{\partial R}+\frac{\partial \Phi}{\partial R} &=& \frac{V^2_{\rm rot}}{R}, \\
\frac{1}{\rho_{\rm g}}\frac{\partial p}{\partial z}+\frac{\partial \Phi}{\partial z} &=& 0,
\end{eqnarray}
with $\rho_{\rm g}$, $p$, $V_{\rm rot}$ and $\Phi$ being the volume
density of the gas, the gas pressure, the azimuthal rotation velocity
(``rotation velocity" in short) and the joint total
potential. Equation (1) describes that the gravitational pull in
radial direction is counterbalanced by the centrifugal force and the
pressure gradient. Equation (2) states that hydrostatic equilibrium
along the symmetry axis, the $z$-direction, is achieved by the balance
between vertical pull of the gravity and the pressure gradient in $z$.

To make the system self-consistent, the Poisson equation must be involved:
\begin{equation}
\nabla^2\Phi = 4\pi G (\rho_{\rm g}+\rho_{\rm DM}+\rho_{\rm s}),
\end{equation}
with $G$, $\rho_{\rm DM}$ and $\rho_{\rm s}$ being the gravitational
constant, and the volume density of dark matter and stars. The total
potential is comprised of the contributions from the dark matter halo,
the stellar disk and the self-gravity of the gas, i.e.,
$\Phi=\Phi_{\rm DM}+\Phi_{\rm s}+\Phi_{\rm g}$. In addition, the ideal
gas law provides the link between the gas density, the gas temperature
and the gas pressure:
\begin{equation}
p=\rho_{\rm g}(\gamma-1)e(T),
\end{equation}
where $\gamma$ represents the ratio of the heat capacities (adiabatic
index), $e$ the specific internal energy and $T$ the gas
temperature. In the case of an ideal gas, the specific internal energy
depends only on temperature, and is given by
\begin{equation}
e={1 \over \gamma-1}{k_{\rm B} T \over \mu m_{\rm p}},
\end{equation}
with $k_{\rm B}$ being the Boltzmann's constant, $\mu$ the atomic
weight and $m_{\rm p}$ the mass of a proton. However, to close the set
of equations, we should either invoke the energy equation or an
equation of state (EoS), which will be used to evolve the system.

A disk which is in hydrodynamic equilibrium should stay in its
original state if we evolve the disk with the same equation of state
which is used to set-up the disk. The numerical results throughout
this paper are based on the isothermal equation of state, i.e.,
\begin{equation}
p=c^2_{\rm s}\rho_{\rm g}, 
\end{equation}
with $c_{\rm s}$ being the sound speed, a temporal and spatial
constant. Equations (1) to (6) then form the basis of our
discussion. In this paper, all the disks are in detailed equilibrium
with the isothermal EoS. If those disks are adopted to evolve with a
cooling function or a polytropic EoS, we can make sure any change in
temperature or dynamics is purely caused by a cooling or a heating
source.

For a polytropic gas, $p=K\rho_{\rm g}^{\Gamma}$. with $\Gamma$ and
$K$ being constant, integrating Eq. (2) gives:
\begin{equation}
\rho_{\rm g}(R,z) = \rho_{\rm g}(R,z=0) \left[ 1-{\Gamma-1 \over c^2_{\rm s}(R,z=0)}\Phi_z(R,z) \right] ^{1 \over \Gamma-1},
\end{equation}
where $\Phi_z(R,z)=\Phi(R,z)-\Phi(R,z=0)$ defines the vertical
potential difference. We have used the fact that $c^2_{\rm s}\equiv
\partial p / \partial \rho_{\rm g}= K\Gamma \rho_{\rm g}^{\Gamma-1}$
when approaching Eq. (7). Note that given $\Gamma \neq 1$, the internal
energy has the following relation:
\begin{equation}
e(T) = \frac{K\rho_{\rm g}^{\Gamma-1}}{\gamma-1}.
\end{equation}
Combining Eq.~(5) and Eq.~(8) gives the temperature field as a
function of position if the gas disk is initialized with a
non-isothermal EoS. As a special case, when $\Gamma \rightarrow 1$,
Eq. (7) then converges to a form for the isothermal gas:
\begin{equation}
\rho_{\rm g}(R,z) = \rho_{\rm g}(R,z=0)\exp{\left(-{\Phi_z{(R,z)} \over c^2_{\rm s}} \right)}.
\end{equation}
As we can see from Eq. (7) and Eq. (9), the vertical structure of gas
disk depends on the gas properties in the midplane and the vertical
potential difference.

To fully characterize a gas disk which is in detailed balance, we need
to specify the velocity, the density and the temperature at every
location in the beginning of the simulation. In the following
sub-sections we study the general properties of the velocity and
density field, which allows us to devise a simple, but effective
method to initialize a 3D gas disk in hydrostatic equilibrium.

\subsection{Azimuthal Rotation Velocity}

In this sub-section, we \revise{treat} the azimuthal rotation velocity
as \revise{generally} as possible. To make the notation concise, we
drop the subscript of gas density, $\rho_{\rm g}$, and restore the
subscript after this sub-section. Without further assumption,
integrating Eq. (2) from $0$ to $z$ gives:
\begin{equation}
\int^z_0 \frac{1}{\rho}\frac{\partial p}{\partial z}{\rm d}z=-\Phi_z(R,z).
\end{equation}
By integrating Eq. (10) in parts, we have:
\begin{equation}
\frac{p(R,z)}{\rho(R,z)}=\left.\frac{p(R)}{\rho(R)}\right|_{z=0}-\int^{z}_{0}\frac{p}{\rho^2}\frac{\partial \rho}{\partial z}{\rm d}z-\Phi_z(R,z).
\end{equation}
Inserting Eq.(11) into (1), we get (see Appendix A for further details):
\begin{eqnarray}
\lefteqn{\frac{V^2_{\rm rot}(R,z)}{R}=\frac{1}{\rho}\frac{\partial p}{\partial R}+\frac{\partial \Phi(R,z)}{\partial R}}\nonumber \\
&=&\left.\frac{1}{\rho}\frac{\partial p}{\partial R}\right|_{z=0}+ \left.\frac{\partial \Phi(R)}{\partial R}\right|_{z=0} \nonumber \\
&-&\int^{z}_{0} \left\lbrace \left(\frac{\partial \rho}{\partial z}\right)\frac{\partial}{\partial R}\left( \frac{p}{\rho^2}\right)-\left(\frac{\partial \rho}{\partial R}\right)\frac{\partial }{\partial z}\left(\frac{p}{\rho^2} \right)\right\rbrace {\rm d}z.
 \end{eqnarray}
Equation (12) shows that the rotation velocity is independent of
height above or below the midplane so long as the integral
vanishes. It is evident that gas with a barotropic equation of state,
i.e., $p(\rho_{\rm g})$, fulfills this requirement. In addition, for
an initially constant temperature ($T$ is everywhere the same
in the beginning) disk, the initial pressure is a function of volume
density only and therefore the integral becomes zero. In this case,
equation (12) can be simplified further:
\begin{equation}
V^2_{\rm rot}(R,z)=R\left.\frac{\partial \Phi}{\partial R}\right|_{z=0}+(\gamma-1)e\left. \frac{\partial \ln \rho}{\partial \ln R}\right|_{z=0}.
\end{equation}
Equation (13) states that the process of specifying the initial
velocity in 3D comes down to knowing the rotation velocity in the
equatorial midplane.
 
\label{subsec:Rotation Velocity}
\label{sec:formulation}

\subsection{Density Distribution}

From now on, to avoid confusion, we restore the subscript for the gas
density. To proceed further, we consider the gas layer to be a very
thin structure embedded in a static potential contributed by the
background spherical dark matter and the stellar disk. Because the gas
disk is observationally thin we neglect the radial variation compared
to the vertical one \revise{(i.e.,
$|({\partial}/{\partial R}(R{\partial \Phi_{\rm g}}/{\partial
R}))/R| \ll |{\partial^2 \Phi_{\rm g}}/{\partial z^2}|$). In
Appendix~E we show that this is a valid assumption for realistic gas
disks}. For an axisymmetric thin disk, the Poisson equation then
reduces to (Binney \& Tremaine 2008):
\begin{equation}
\frac{{\rm d}^2 \Phi_{\rm g}}{{\rm d} z^2}=4\pi G \rho_{\rm g}.
\end{equation}
with $\Phi_{\rm g}$ being the potentials contributed by the gas. In
the following, we focus only on disks with initially constant
temperature, i.e., the rotation velocity required for equilibrium has
no dependence on the height above or below the midplane.

\subsubsection{Density Method}

Differentiation Eq.~(2) with respect to $z$ and inserting Eq.~(14)
leads to a second-order non-linear differential equation:
\begin{equation}
\frac{{\rm d}^2 p}{{\rm d} z^2}-\frac{1}{\rho_{\rm g}}\frac{{\rm d} \rho_{\rm g}}{{\rm d} z}\frac{{\rm d} p}{{\rm d} z}+\rho_{\rm g}(4\pi G\rho_{\rm g}+\frac{{\rm d}^2 \Phi_{\rm s}}{{\rm d} z^2} + \frac{{\rm d}^2 \Phi_{\rm DM}}{{\rm d} z^2})=0,
\end{equation}
\revise{with $\Phi_{\rm s}$ and $\Phi_{\rm DM}$ being the potentials
contributed by the stellar disk and the dark matter, respectively}. So
far, Eq. (15) is still general with respect to any kind of equation of
state. However, a single equation with two unknowns $p$ and $\rho_{\rm
g}$ is not solvable. To continue with Eq. (15), in this sub-section,
we assume that the gas is barotropic, i.e., $p(\rho_{\rm g})$. Given
density distributions of stars, the dark matter and the boundary
conditions in the midplane:
\begin{equation}
\rho_{\rm g}(R,0)=\rho_0(R)  \textrm{ and } \frac{{\rm d} \rho_{\rm g}}{{\rm d} z}=0, 
\end{equation}
equation (15) can be solved by numerical integration, e.g., using the
Runge-Kutta method.  For a single-component, self-gravitating, locally
isothermal disk ($c_{\rm s}(R)$ can be a function of radius), Eq. (15)
has an exact solution:
\begin{equation}
\rho_{\rm g}(R,z)=\rho_0(R)\textrm{sech}^2 (z/h),
\end{equation}
with $\rho_0(R)$ being the gas volume density in the midplane,
$h=\sqrt{c^2_{\rm s}/2\pi G\rho_0}$ the scale-height and $c_{\rm s}$
the local isothermal sound speed. According to Eq. (17) and since the
midplane volume density, $\rho_0(R)$, generally decreases with radius,
to keep the scale-height a constant, the sound speed and therefore the
temperature must be a function of radius.

Equation (15) is the simplified version of the formula derived by NJ02 (see also
Kim \etal 2002a), where they investigated the vertical structure in a
gravitationally coupled, multi-component galactic disk. It is
important to notice that all calculations can be done locally without
the need of exchanging information between processors and therefore
greatly reduces the complexity of coding.

In principle Eq. (15) allows one to compute the density of the gas
such that the disk initially is in hydrostatic equilibrium. The actual
implementation using Eq. (15) does not guarantee the positivity of the
density. In particular, at large radii $\rho_{\rm g}(R,z)$ is
typically close to zero, and small errors due to the numerical
integration often yield negative densities. This problem is especially
relevant when using the adaptive-mesh refinement techniques.

Initializing a gas disk with AMR usually starts with the coarsest
grid. A natural selection of the integration step is the cell
size. Then a problem immediately rises when solving Eq. (15) to
specify the volume density.  Supposing that the cell size is much
larger than the scale-height of the gas disk, the errors introduced by
the coarse integration may lead to negative densities on the outskirt
of the computation domain. One might think the integration can be done
by using either adaptive integration intervals or simply a fixed
integration interval which is much smaller than the cell
size. However, the improvements only guarantee the convergence of the
solution not the positivity. Nevertheless, because of the generality
of Eq.~(15), density method is still valuable for semi-analytic
research.

\label{subsubsec: Density Method}
\subsubsection{Potential Method}

In this sub-section, we develop another route for specifying the
density distribution. We stress that the following derivation is only
applicable to initially isothermal disks. With this constraint,
integrating Eq. (2) gives:
\begin{equation}
\rho_{\rm g}(R,z)=\rho_0(R)  \exp{\left(-\frac{\Phi_z(R,z)}{(\gamma-1)e}\right)}.
\end{equation}
Combining Eq. (14) and Eq. (18), a second-order non-linear equation
for the vertical potential difference of gas is obtained:
\begin{equation}
\frac{{\rm d}^2\Phi_{{\rm g},z}}{{\rm d}z^2}=4\pi G\rho_0(R)  \exp{\left(-\frac{\Phi_z(R,z)}{(\gamma-1)e}\right)}.
\end{equation}
Given the analytic forms of $\Phi_{\rm DM}$ and $\Phi_{\rm s}$ the
only unknown is the potential difference of gas, $\Phi_{{\rm
g},z}=\Phi_{\rm g}(R,z)-\Phi_{\rm g}(R,z=0)$. Similar to the density
method, given the boundary conditions $\rho_0(R)$, $\Phi(R,z=0)$ and
${\rm d}\Phi(R,z=0)/{\rm d}z=0$, numerical integration can be applied
to solve Eq. (19). By inserting $\Phi_z$ obtained by integrating
Eq. (19) into Eq. (18), the density distribution is acquired. Notice
that what really matters to us is the potential difference, not the
absolute value. This means the value of $\Phi_{\rm g}(R,z=0)$ can be
an arbitrary constant, although we do know the values of $\Phi_{\rm
DM}(R,z=0)$ and $\Phi_{\rm s}(R,z=0)$.

The merit of this formulation is evident, the occurrence of negative
density is avoided by Eq. (18). Tiny errors in the potential
difference will not do any harm to the positivity of the gas
density. Numerical experiments show that in normal cases in which both
the density method and potential method work, the solutions are
consistent.

At a given radius, $R$, solving Eq. (19) only provides us with information
about the potential difference, $\Phi_z(R,z)$. However, a useful
byproduct of the potential method is that it is possible to acquire a
good approximation of the total potential by:
\begin{equation}
\Phi_{\rm g}(R,z)=\Phi_{\rm g}(R,z=0)+\Phi_{{\rm g},z}(R,z),
\end{equation}
as long as we know the potential in the midplane, $\Phi_{\rm
g}(R,z=0)$. Equation~(20) is an approximation because the use of
Eq.~(19) is based on the reduced Poisson equation Eq.~(14) in which
the variation in radial direction is ignored. The gradient of the
potential $\Phi_{\rm g}(R,z=0)$ determines the velocity field required
while the vertical potential difference $\Phi_{{\rm g},z}(R,z)$ gives
the vertical structure of the disk. In principle, the radial force,
which is associated with $\Phi_{\rm g}(R,z=0)$, in the equatorial
plane for an axially symmetric density distribution can be evaluated
precisely by the equation (A.17) in Casertano (1983). This allows us to
obtain the total potential without fully solving the Poisson
equation. In practice, if the initialization is performed with
multi-node clusters, each node only keeps part of the information
about the density distribution, data exchange with AMR itself is
technically challenging. In Section 3, for an exponential disk, the
numerical results will show that the use of Eq. (29) is a good
approximation for most of our interests. The corresponding $\Phi_{\rm
g}(R,z=0)$ associated with Eq. (29) can be found in the book by Binney
\& Tremaine (2008), Eq. (1.164a).

Equation (20) is useful, because involving the total potential into
the formulation is an important step for self-consistently building up
the combined disks comprised of a live stellar disk and a gas
disk. Extension to the work of Shu (1969), Kuijken \& Dubinski (1995,
hereafter, KD95) develop a self-consistent disk-bulge-halo model for
galaxies. The distribution function built by Eq. (6) in KD95 involves
the potential differences $\Phi_z$ and $\Phi(R,0)-\Phi(R_{\rm c},0)$,
with $R_{\rm c}$ \revise{the radius of the guiding center}. The
potential method presented here can be naturally incorporated into the
framework of KD95. Therefore, in this paper, all the disks are
initialized by the potential method.

\label{subsubsec: Potential Method}

\subsubsection{Exponential Disk}

Some studies have assumed that the midplane density of a 3D gas disk
has an exponential form (Tasker 2006, Agertz 2009). However, as we now
demonstrate, in general this results in a surface density distribution
that peaks at a specific non-zero radius, giving rise to a ring-like
feature. We assume a gas disk with the popular $\textrm{sech}^2$
vertical profile:
\begin{equation}
\rho_{\rm g}(R,z)=\rho_{\rm c} \exp(-R/R_{\rm d}) \textrm{sech}^2\left(\frac{z}{h(R)}\right),
\end{equation}
with $\rho_{\rm c}$ being the central volume density, $R_{\rm d}$ the
disk scale-length and $h(R)$ the scale-height as a function of
radius. The surface density then reads:
\begin{equation}
\Sigma(R)=\int^{\infty}_{-\infty}\rho_{\rm g}(R,z){\rm d}z=2\rho_{\rm c}\exp(-R/R_{\rm d})h(R).
\end{equation}
Based on Eq. (22), we measure the scale-height of a disk at certain radius
by $h(R)=\Sigma(R)/(2\rho_0(R))$. The extrema of the surface density
can be evaluated by taking the derivative to Eq. (22):
\begin{equation}
\frac{{\rm d}\Sigma(R)}{{\rm d}R}=2\rho_{\rm c}\exp(-R/R_{\rm d})\left(\frac{{\rm d}h(R)}{{\rm d}R}-\frac{h(R)}{R_{\rm d}}\right)=0.
\end{equation}
We Suppose that the disk is linearly flaring, i.e., $h(R)=h_0 +
R/R_{\rm h}$, with $h_0$ being the minimum scale-height of the disk
and $R_{\rm h}$ a factor controlling the degree of flaring. The peak
of the surface density then locates at $R_{\rm peak}=R_{\rm
d}-h_0R_{\rm h}$. Whenever the $R_{\rm peak}$ is positive, we get a
ring in surface density. However, a ring in the surface gas density is
not commonly seen in a real disk galaxy. An exponential profile in the
total gas is prevalent in disk galaxies (Leroy \etal 2008).

In order to avoid this feature, it is advantageous to specify the
actual surface density of the disk, rather than its midplane
density. In the case of the exponential profile, the surface density
reads:
\begin{equation}
\Sigma(R)=\Sigma_0\exp{(-R/R_{\rm d})}=\int^{\infty}_{-\infty}\rho_{\rm g}(R,z){\rm d}z,
\end{equation}
with $\Sigma_0$ being the surface density in the galactic
centre. Combining Eq. (24) and Eq. (18), the volume density in the
midplane can be expressed as:
\begin{equation}
\rho_0(R)=\frac{\Sigma_0\exp{(-R/R_{\rm d})}}{\int^{\infty}_{-\infty}\exp{\left(-\Phi_z/[(\gamma-1)e]\right)}{\rm d}z}.
\end{equation}
It shows that the correct volume density in the midplane for the
desired surface density profile can be obtained iteratively. Given a
initial guess for $\rho_0(R)$, $\Phi_z$ is evaluated via Eq. (19) and
also the integral appears in Eq. (25). One needs to iterate between 
Eq. (19) and Eq. (25). However, depending on the quality of the initial 
guess, convergence can be reached very fast. For instance, with the 
initial guess being $\rho_0(R)=\Sigma_0 \exp(-R/R_{\rm d})$, a six-time 
iteration already gives us a reasonable exponential disk.

We pursue the exponential disk for several reasons. One is simply
because it is commonly seen in disk galaxies. Another is that we have
a better control of the total mass. As we can see, if we specify the
midplane volume density instead of the surface density, we do not
exactly know the total mass until we finish the integration. Without
knowing the total mass in advance, evaluating the circular velocity
contributed by the self-gravity will not be a trivial
task. Nevertheless, in principle, any profile of the surface density
can be achieved simply by the process introduced in this sub-section.
\label{subsubsec: Exponential Disk}

\label{subsec:Density Distribution}

\section{Implementation and Tests}

\subsection{Simulation Parameters}
\begin{table}
\label{tab:cases}
\caption{Models' Parameters.}
\begin{tabular}{cccl}
\hline
Run  & $T$ (K) & $M_{\rm s} ({\rm M}_{\odot})$ & Figure\\
\hline\hline
Gas0 & $4\times 10^4$ & - & (1),(2),(3)\\
Gas1 & $2\times 10^4$ & - & (4),(5)\\
Gas2 & $1\times 10^4$ & - & (4),(5),(8)\\
Gas3 & $9\times 10^3$ & - & (4),(5)\\
Gas4 & $8\times 10^3$ & - & (4),(5)\\
GasStar1 & $7\times 10^3$ & $4\times 10^{10}$ & (6),(7),(8)\\
GasStar2 & $6\times 10^3$ & $4\times 10^{10}$ & (6),(7)\\
GasStar3 & $5\times 10^3$ & $4\times 10^{10}$ & (6),(7)\\
GasStar4 & $4\times 10^3$ & $4\times 10^{10}$ & (6),(7)\\
\hline
\end{tabular}
\begin{minipage}[c]{0.5\textwidth}
*All disks have a gas mass of $10^{10} M_{\odot}$.
\end{minipage}
\end{table}
In this Section, we test the ideas outlined in the previous
Section. We implement the method in the AMR-code RAMSES (Teyssier
2002). RAMSES uses grid-based Riemann-solvers for the
magneto-hydrodynamics (MHD) and particle-mesh (PM) technique for the
collisionless physics. It has a fully parallelized Poisson solver 
with periodic boundary conditions, which we use for this paper. 
Gas disks which are initialized isothermally with an exponential 
surface density of a scale-length of 3.5 kpc and a total mass of 
$10^{10} M_{\odot}$ are embedded in a static potential. An isothermal 
equation of state is used to evolve the disks throughout this paper.

The tests are mainly divided into two groups, one group is evolved
with a static stellar potential (models with the prefix GasStar), the
other without (models with the prefix Gas). Gas1 to Gas4 are M33-like
gas-rich galaxies, while GasStar1 to GasStar4 are more similar to the
Milky-Way. The main parameters of the models are listed in Table
1. The size of the computational domain is 250 kpc on a side. Up to 12
levels of refinement are used for those runs without stellar
potential, and 13 levels for the other group, i.e., the corresponding
highest spatial resolutions are about 60 pc and 30 pc, respectively.

The volume density of the halo is described by the NFW profile
(Navarro, Frenk \& White 1997):
\begin{equation}
\rho_{\rm DM}(r)=\frac{M_{200}}{4\pi
f(c)r_{200}}\frac{cx}{r^2(1+x)^2},
\end{equation}
with the Virial mass $M_{200}=10^{12}{\rm M}_{\odot}$, $x=rc/r_{200}$,
concentration parameter $c=12$, distance $r=\sqrt{R^2+z^2}$, Virial
radius $r_{200}=213$ kpc and $f(c)=\ln(1+c)-c/(1+c)$. The Virial
radius ($r_{200}$) is a radius within which the averaged matter
density is 200 times the critical density.

The density distribution of the stellar disk reads (Miyamoto \& Nagai
1975, Binney \& Tremaine 2008):
\begin{equation}
%\Phi_s(R,z)=-\frac{GM_s}{\left(R^2+\left[a+(z^2+b^2)^{1/2}\right]^2 \right)^{1/2}},
\rho_{\rm s}(r)=\left(\frac{b^2M_{\rm s}}{4\pi}\right)\frac{aR^2+(a+3\sqrt{z^2+b^2})(a+\sqrt{z^2+b^2})^2}{\left[ R^2+(a+\sqrt{z^2+b^2})^2\right]^{5/2}(z^2+b^2)^{3/2}},
\end{equation}
with $M_{\rm s}=4\times10^{10}{\rm M}_{\odot}$ being the mass of the
stellar disk, $a=3.5$ kpc and $b=0.2$ kpc the shape parameters.

In light of the result drawn from Section 2.1, for an initially
constant temperature setup, we only need to know the circular velocity
in the midplane for initializing the velocity field. The rotation
velocity, $V_{\rm rot}$, is decomposed into four components:
\begin{equation}
 V^{2}_{\rm rot}=V^{2}_{\rm DM}+V^2_{\rm s}+V^2_{\rm g}+V^2_{\rm p},
\end{equation}
where $V_{\rm DM}$, $V_{\rm s}$, $V_{\rm g}$ are the circular
velocities corresponding to the dark matter halo, the stellar disk and
the gas disk, and $V_{\rm p}$ denotes the contribution due to the
pressure gradient.

In this paper, we have the analytic form for $V_{\rm DM}$ and $V_{\rm
s}$. For the contribution from the gas disk and pressure gradient, we
take the approximation for an infinitesimally thin disk with
exponential surface density as described in Eq. (24). We set:
\begin{eqnarray}
V^2_{\rm g}(R)&=& 4\pi G\Sigma_0 R_{\rm d}
y^2[I_0(y)K_0(y)-I_1(y)K_1(y)] \\ V^2_{\rm p}(R) &=&
(\gamma-1)e\left.\frac{\partial \ln \rho}{\partial \ln R} \right
|_{z=0},
\end{eqnarray}
with $y=R/(2R_{\rm d})$, $I_0$, $K_0$, $I_1$ and $K_1$ being the
modified Bessel functions of the first and second kinds of
zeroth/first-order, respectively. Equation (30) derives from the
second term of Eq. (13). However, contribution from pressure gradient
in the midplane can only be evaluated after the gas disk is set
up. Note that for an exponential disk, surface and volume densities
decrease with radius and hence $V^2_{\rm p}$ is negative.

\subsection{A Stable Disk}
%%% fig. 1
\begin{figure*}
{
\centerline{\psfig{figure=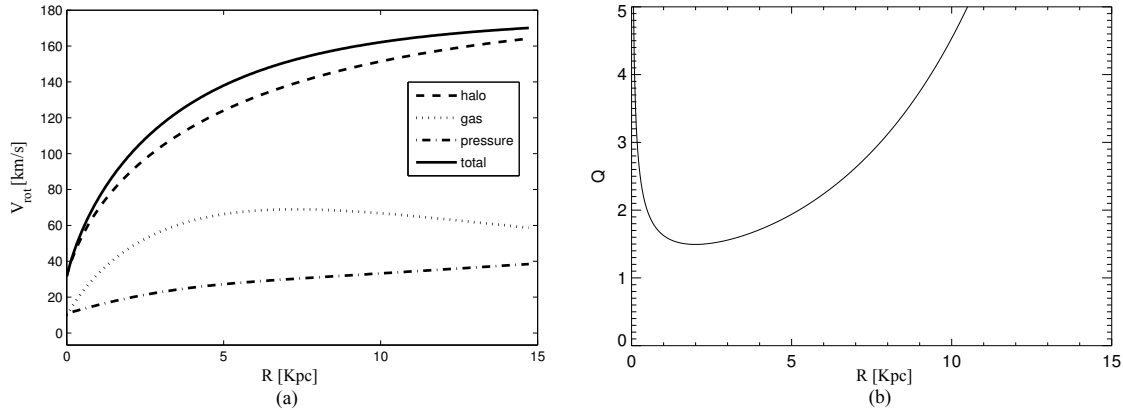,width=0.85\textwidth}}
}
\caption{Model Gas0: (a) The total rotation velocity (solid) and
contributions from dark matter halo (dashed), gas (dotted) and gas
pressure (dash-dotted). Note that we plot the absolute value of the 
pressure gradient to have positive values for the direct comparison. 
It should be in opposite sense to the gravity. In this model contributions 
from the gas self-gravity and the pressure gradient is not negligible. 
(b) The $Q$ value of model Gas0 as a function of radius as defined by 
Eq. (32). The $Q$ is well above the threshold value $Q_{\rm th}=1$, 
thus the disk is expected to be stable. No structure should develop with time.}
\label{fig:galprop}
\end{figure*}
%
 
%%% fig.2
\begin{figure*}
{
\centerline{\psfig{figure=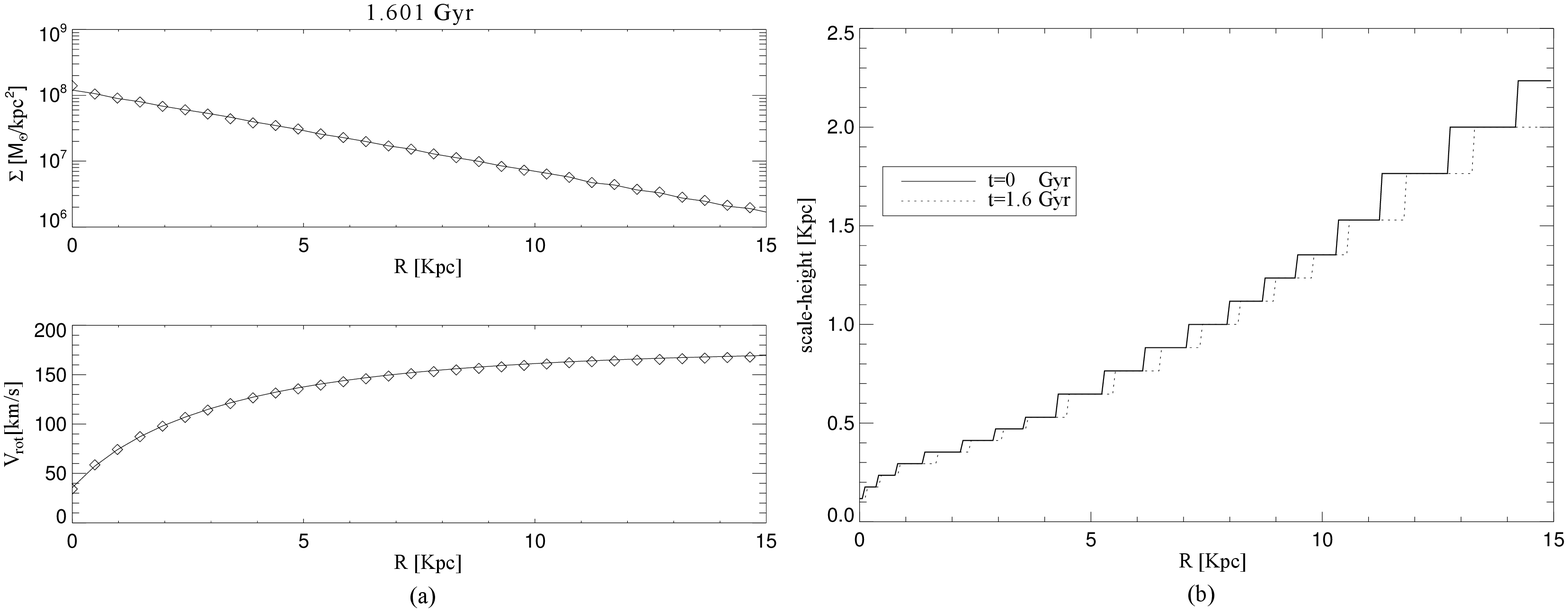,width=0.85\textwidth}}
}
\caption{Model Gas0: The evolution of 1.6 Gyrs of a stable disk. (a)
  The evolution of the surface density (upper panel) and the rotation
  curve (lower panel) at $t=0$ Gyr (solid) and $t=1.6$ Gyr
  (diamond). Overall, the surface density and rotation curve are kept
  very well over 4 orbital periods. (b) The evolution of the
  scale-height at $t=0$ Gyr (solid) and $t=1.6$ Gyr (dotted). The
  small change in scale-height indicates that the required circular
  velocity is overestimated probably due to the approximation of
  Eq. (14) and Eq. (30). In all, the disk still stays well in the
  initial condition. \revise{The step-wise character of the
  scale-height reflects our discretization and the change of spatial
  resolution due to the AMR.}}
\label{fig:galprop}
\end{figure*}
%

%%% fig.3
\begin{figure*}
{
\centerline{\psfig{figure=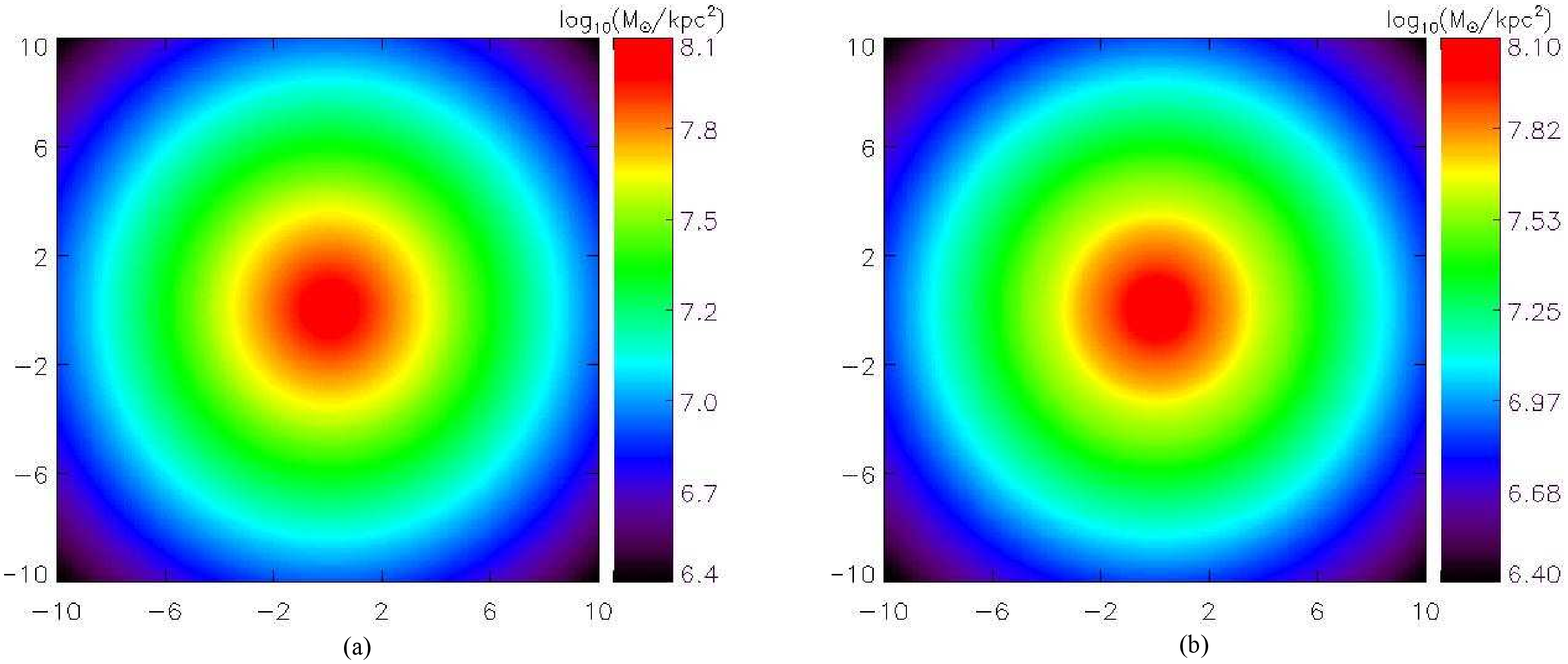,width=0.85\textwidth}}
}
\caption{Model Gas0: The size of the images is 20 kpc $\times$ 20
kpc. The evolution of the surface density at (a) $t=0$ Gyr and (b) at
$t=1.6$ Gyr. This figure shows that no structure is developing over
secular time-scale.}
\label{fig:galprop}
\end{figure*} 

To demonstrate that the disk built by the potential method described
in Section 2 is in detailed equilibrium, we start with a stable
equilibrium disk in model Gas0. In this test, the stellar disk is
deliberately removed. Without the dynamical support from the stellar
disk, the self-gravity of the gas plays the dominant role in
determining the vertical structure of the disk and provides a not
negligible contribution to the rotation velocity.

Figure 1a decomposes the rotation curve into the different
contributions by the halo, the gas and the pressure gradient. Note
that the forces of the self-gravity and the pressure gradient are in
opposite sense, the self-gravity pulls matter inwards while the
pressure gradient pushes outwards. In this figure, $V_{\rm p}$ is
shown in its absolute value. If we ignore the pressure gradient, the
disk would rotate too fast and gradually drift outward. Figure 1b
shows the conventional Toomre's Q defined by:
\begin{equation}
Q=\frac{c_{\rm s} \kappa}{\pi G \Sigma_{\rm g}}.
\end{equation}
with $\kappa$ being the epicyclic frequency. It shows that the $Q$ is
well above $Q_{\rm th}=1$, the threshold value for stability, at all
radii. The disk is hot enough to keep the disk stable and no structure
should develop with time.

We let the disk evolve for 1.6 Gyrs (four orbits for the gas at 10
kpc) and check how well the disk properties are kept. Figure 2a
presents the evolution of the surface density and the rotation
curve. The solid lines represent the initial states and the diamond
symbols are the status after an evolution of 1.6 Gyrs. The surface
density is obtained by projecting along the symmetry axis and the
rotation curve is evaluated by the mass-weighted circular velocity,
$\bar{v}_{\rm rot}(R)=\int \rho_{\rm g}(R,z)v_{\rm rot}(R,z){\rm
d}z/\Sigma_{\rm g}(R)$. Although a small amount of mass accretes onto
the very central part of the disk, overall the surface density and the
rotation curve are kept very well. Mass accretion into the center
seems unavoidable for a Cartesian-grid code mainly because too small a
number of cubic cells is used to mimic the circular motion in the
centre. This accretion will be eventually halted by the pressure
gradient built by the accumulating material.

Figure 2b shows the evolution of the scale-height. The solid line
represents the initial state and the dotted line the evolution after
1.6 Gyrs. Upon closer inspection we find that the disk undergoes a
very small amount of mass-redistribution in the radial direction,
which we believe to be a consequence of our two approximations when
initializing the disk. One is from the reduced Poisson equation,
Eq. (14), and the other is from the use of Eq. (29). Equation (29)
overestimates the circular velocity needed to support the disk. The
thickness of the disk reduces the potential in the midplane by a 
few percent (see appendix B). Figure 3 shows the snapshots of the 
face-on surface density at $t=0$ (Fig. 3a) and at $t=1.6$ Gyr 
(Fig. 3b). No structure is developing during the course of the simulation.

To sum up, figures 1 to 3 indicate that without external perturbation
the disk is quiet over secular time-scales. The shape of such a disk is
naturally flaring, i.e., the scale-height increases with radius. The
ideas described in Section 2 are able to treat the initial condition
self-consistently. A well-balanced disk is especially useful to probe
the onset of disk instability as described in the next Section.

\section{Axisymmetric Instability}

The question of disk stability has been investigated for more than four decades since
the pioneering works by Toomre (1964) for collisionless stars and
\revise{Goldreich \& Lynden-Bell (1965) for gas sheets}. Understanding
the origin and evolution of disk structure is challenging. If the disk
is stable like our model Gas0, no structures can form. On the other
hand, if the disk is highly unstable, the surface density will quickly
fragment and develop a clumpy and chaotic-looking appearance. There
will be no well-organized structures. The striking spiral appearance of 
many nearby disk galaxies indicates that those disks are marginally stable.

For an infinitesimally thin disk, the instability threshold is at
$Q_{\rm th}=1$ (Toomre 1964). The first theoretical work to include
the finite thickness of a self-gravitating gas disk is that by
Goldreich \& Lynden-Bell (1965). Some authors have investigated the
stability of finite thickness gas disks in numerical simulations (both
in 2D and 3D) using local patches within a shearing box (Kim \&
Ostriker 2006, 2002a; Gammie 2001). This technique, in 2D, has also
been used by Kim \& Ostriker (2007) to investigate the interaction
between the gas disk and a live stellar disk. Shetty \& Ostriker
(2006) used global 2D simulations in which they incorporated the
effect of finite disk thickness by diluting the gravitational
force. For 3D global disk calculations, see Li, Mac Low \& Klessen 
(2005a, 2005b, 2006), who investigate the relation between disk 
instability and star formation rate. These studies all agree that 
although the inclusion of the thickness does not have a qualitative 
impact on the disk instability, it does shift the threshold value 
of instability quantitatively. In addition, accounting for disk 
thickness may have a large impact on the evolution of a disk, such 
as the development of spurs or the wiggle instability 
(Kim \& Ostriker 2002b, 2006).

In this paper, armed with a well-balanced gas disk, we revisit the
axisymmetric instability of disks in 3D global fashion. We first
derive the reduction factor $F$ which reflects the reduction of the
gravity due to the finite thickness of the disk. Then the
corresponding instability threshold $Q_{\rm th}(R,T)$ derived from a
semi-analytic calculation is compared with the numerical results. In
the final sub-section, we also explore the impact of the presence of
a static stellar potential on the axisymmetric instability.

\subsection{The impact of thickness on disk stability}

The Fourier component of the perturbed gravitational potential,
$\Phi_k$, of an infinitesimally thin disk is given by:
\begin{equation}
\Phi_k=-\frac{2\pi G\Sigma_k}{|k|}e^{ikx-k|z|},
\end{equation}
where $k$ represents the wave number of the Fourier components and
$x=R-R_0$ being the radial deviation for an axisymmetric
perturbation. Supposing that a 3D disk is piled up by a stack of
infinitesimally thin gas layers, we approximate the effect of the disk
thickness by superimposing the contribution from every razor-thin
layer:
\begin{equation}
\Phi_k(z) = -\frac{2\pi G\Sigma_ke^{ikx}}{|k|}\int^{\infty}_{-\infty}e^{-k|z-h|} \frac{\textrm{sech}^2(h/h_z)}{2h_z}{\rm d}h,
\end{equation}
with $h_z$ being the scale-height of the disk. In Eq. (33), we model
the vertical structure of the gas disk by a $\textrm{sech}^2$
function. This is valid especially for the inner part of disks where
the vertical structure is mainly determined by the self-gravity of the
gas. See also the Fig. D1 in Appendix D. Equation (33) leads to the
Fourier potential in the midplane:
\begin{equation}
\Phi_k(z=0) = -\frac{2\pi G\Sigma_ke^{ikx}}{|k|} F(k,h_z),
\end{equation}
with $F(k,h_z)$ being the reduction factor described by (see Appendix C):
\begin{equation}
F(k,h_z)=1-\frac{1}{2}kh_z\left[ H\left(\frac{kh_z}{4}\right)-H\left(\frac{kh_z}{4}-\frac{1}{2}\right)\right],
\end{equation}
with $H$ being the harmonic number defined by:
\begin{equation}
H(\alpha) = \int^{1}_{0}\frac{1-y^{\alpha}}{1-y}{\rm d}y.
\end{equation}
The Lin-Shu (1964) dispersion relation for the axisymmetric
perturbation is then modified to:
\begin{equation}
\omega^2=\kappa^2-2\pi G\Sigma_0 |k|F(k,h_z) + c^2_{\rm s}k^2.
\end{equation}
The dispersion relation states that on small scales ($k\rightarrow
\infty$) the disk is stabilized by gas pressure, i.e., the term
$c^2_{\rm s}k^2$. Large scales ($k\rightarrow 0$) are regulated by
global shear, i.e., the $\kappa^2$ term. The instability however
happens at intermediate wavelengths, much smaller than the disk size
but still larger than the thickness of the disk. In this region,
neither global shear nor gas pressure can resist the
gravitational collapse. The reduction factor, $0<F\leq 1$, softens the
effect of self-gravity and makes the disk more stable.

%We probe the threshold value of $Q_c$=max($Q_{thick}(R,k)$) by searching for the maximum of:
%
%\begin{equation}
%Q_{thick} = 2\frac{|k|c_s}{\kappa}F(k,h_z)/(1+\frac{|k|^2 c^2_s}{\kappa^2}).
%\end{equation}
%
Given a certain radius $R$ and temperature $T$, we obtain the
threshold value $Q_{\rm th}(T,R)$ by probing the maximum value along
the neutral curve defined by setting $\omega^2=0$ in Eq. (37) and
calculating the epicyclic frequency, $\kappa$, from the rotation
curve. Similar to the conventional Toomre criterion for the stability
of an infinitesimally thin disk, $Q_{\rm th}$ is a threshold curve for
thick disks. Above $Q_{\rm th}$ the disk is stable and otherwise
unstable. Since the $Q_{\rm th}$ is a function of both temperature and
radius, it is convenient to define the critical value $Q_{\rm crit}$,
which is the value of $Q_{\rm th}$ for which $Q_{\rm th}(T,R)/Q(R)=1$,
and the corresponding critical temperature $T_{\rm crit}$.

The solid lines shown in Fig. 4 represent the threshold value $Q_{\rm
th}$ as a function of radius. Each plot corresponds to a disk of
different temperature. The dash-dotted lines are the actual $Q$ values
defined by Eq. (31) of the different models. From these figures, the
most unstable radius is around $R=2$ kpc. The corresponding surface
densities after an evolution of 750 Myrs are shown in Fig. 5. The gas
at the most unstable region has revolved for more than four orbital
periods around the disk center.

These figures shows that the prediction of $Q_{\rm crit}$ and the
numerical results match quite well. The Q value of Gas1 is well above
the solid line and shows a featureless surface density. As shown in
Gas2 and Gas3, with the decrease in temperature, the $Q_{\rm th}$
curves shift up and the disks' $Q$ curves come down. As a consequence, 
the disk starts to develop multi-armed structure, which is very likely 
caused by swing amplification, as discussed in Section 5. And finally in Gas4,
the curves $Q$ and $Q_{\rm th}$ intersect. The disk fragments and
starts to behave chaotically. A more detail calculation shows that the
two curves just touch each other at a temperature $T_{\rm
crit}=8.5\times 10^3 {\rm K}$ with the maximum threshold $Q_{\rm
crit}=0.693$, which is close to $Q_{\rm crit}=0.676$ of Goldreich \&
Lynden-Bell's (1965) analysis but away from the numerical result,
$Q_{\rm crit}=0.647$, of Kim \etal (2002a). However, the actual value
of $Q_{\rm crit}$ is model dependent. Different models of the dark
matter, the stellar disk and even the EoS will all affect the
resulting value of $Q_{\rm crit}$.

%%%%%% figure 4
\begin{figure*}
{
\centerline{\psfig{figure=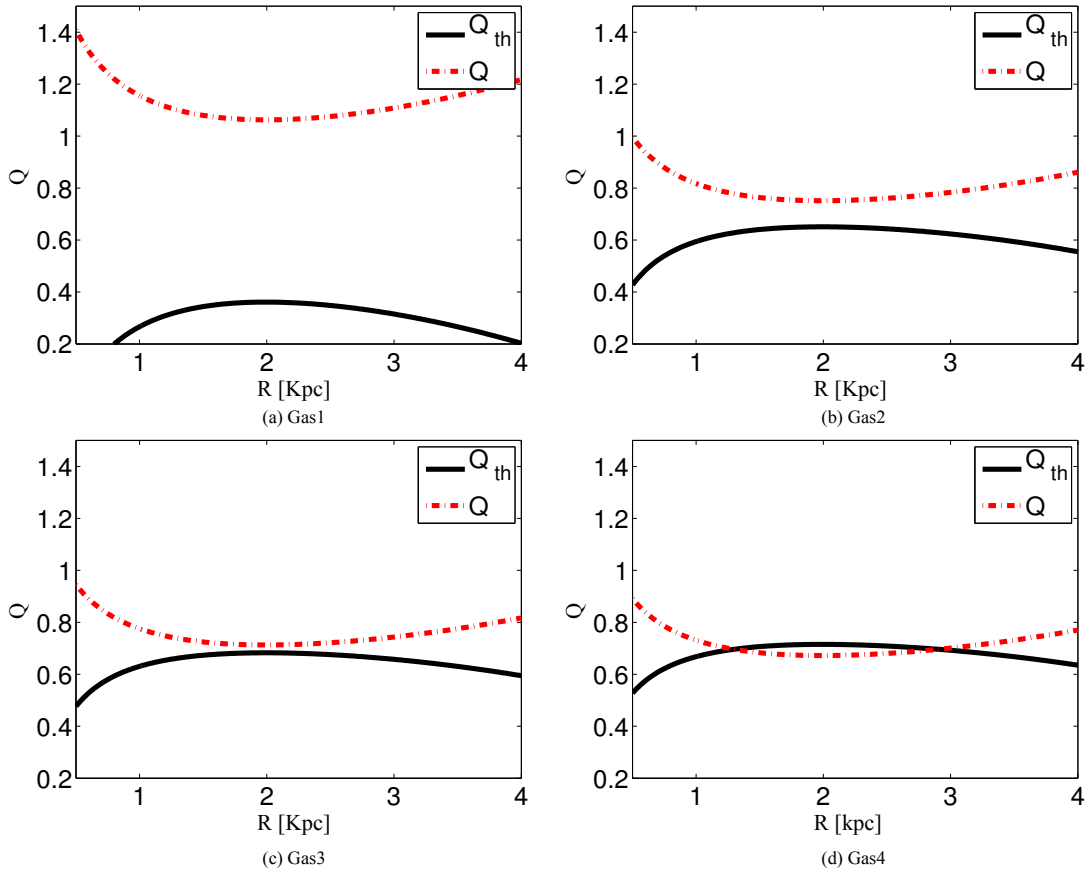,width=0.85\textwidth}}
}
\caption{Plots (a) to (d) correspond to models from Gas1 to Gas4,
respectively. In each plot, curves of the disk $Q$ (dash-dotted) and
the threshold value $Q_{\rm th}$ (solid) are put together to probe the
onset of axisymmetric instability. $Q_{\rm th}(R)$ is a obtained by
probing the maximum value along the neutral curve for a given
radius. Information of the disk thickness has been encapsulated in the
reduction factor defined by Eq. (36). When the two curves meet, we
expect the disk fragments very fast. This figure shows that the most
unstable region is about the radius $R=2$ kpc. The fact that the
$Q_{\rm th}$ curves are well below unity shows the impact of the disk
thickness on the disk stability.}
\label{fig:galprop}
\end{figure*}

%%% figure 5

\begin{figure*}
{
\centerline{\psfig{figure=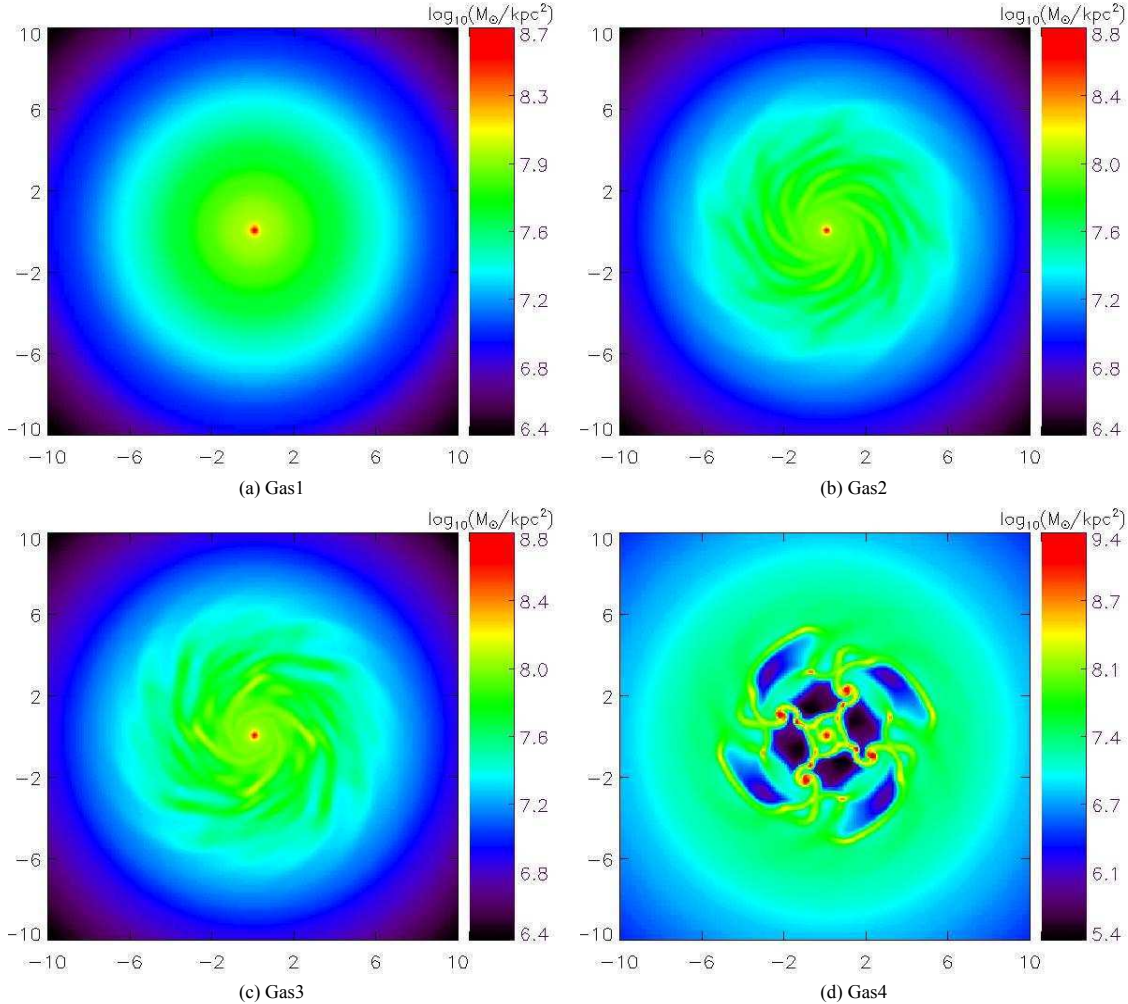,width=0.85\textwidth}}
}
\caption{Images (a) to (d) correspond to models Gas1 to Gas4,
respectively. They show the face-on surface density at $t=750$ Myr. 
The size of the images are 20 kpc $\times$ 20 kpc. The gas at the 
most unstable radius has orbited around the center for more than 
four times. (a) Since the disk $Q$ is well above the threshold value 
$Q_{\rm th}$, the disk is featureless. In models Gas2(b) and Gas3(c) 
the disk $Q$ is approaching $Q_{\rm th}$ around $R=2$ kpc, both 
disks are developing self-induced spirals due to swing amplification. 
(d) The disk fragments very fast once $Q$ and $Q_{\rm th}$ intersect.}
\label{fig:galprop}
\end{figure*}

\subsection{The inclusion of stellar potentials}
%%%% figure 6
\begin{figure*}
{
\centerline{\psfig{figure=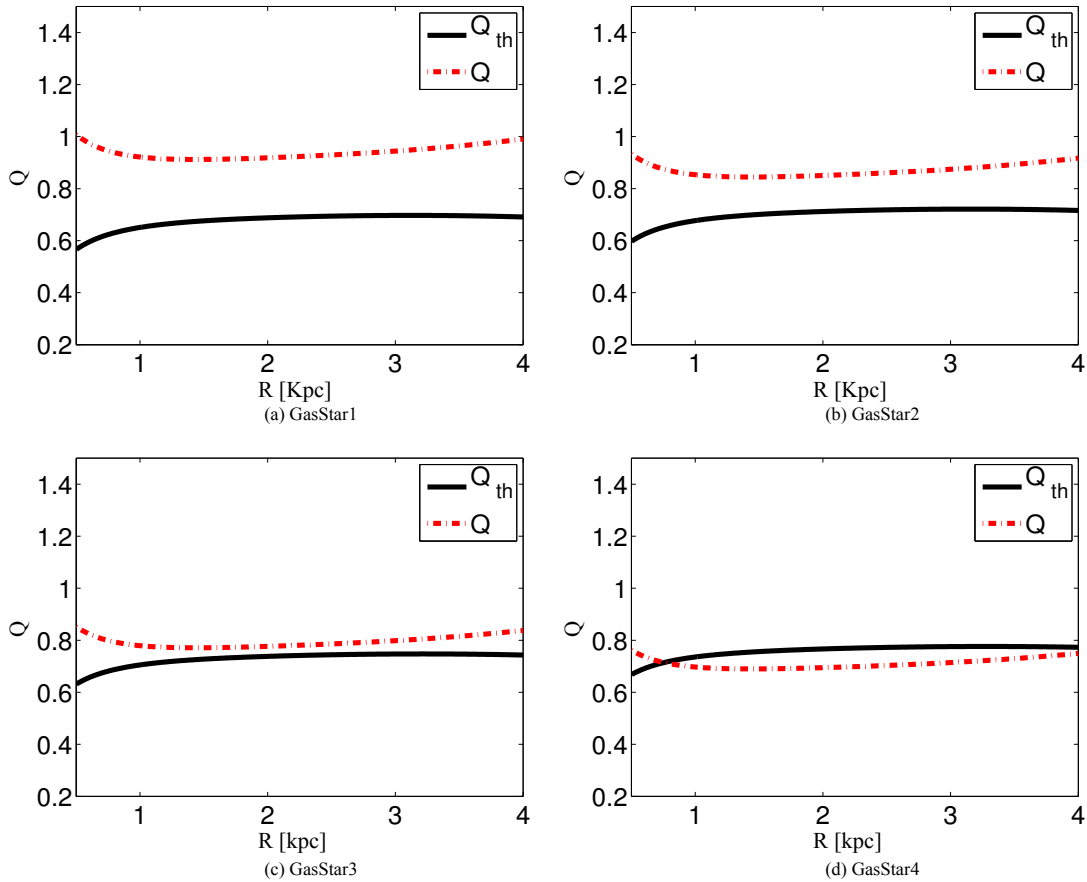,width=0.85\textwidth}}
}
\caption{Plots (a) to (d) correspond to models GasStar1 to GasStar4,
respectively.: The $Q$ (dash-dotted) and $Q_{\rm th}$ (solid) curves
of the gas disks of different temperatures. The presence the stellar
potential stabilizes the disks through changing the rotation curve and
destabilizing the disk by increasing local gravitational force. The
effect of disk thickness is included via the reduction factor
Eq. (36). We need to lower the temperature down to T$=7\times 10^3$ K
in order to probe the onset of axisymmetric instability. Overall, the
presence of the stellar potential stabilizes the disk.}
\label{fig:galprop}
\end{figure*}

%%% figure 7
\begin{figure*}
{
\centerline{\psfig{figure=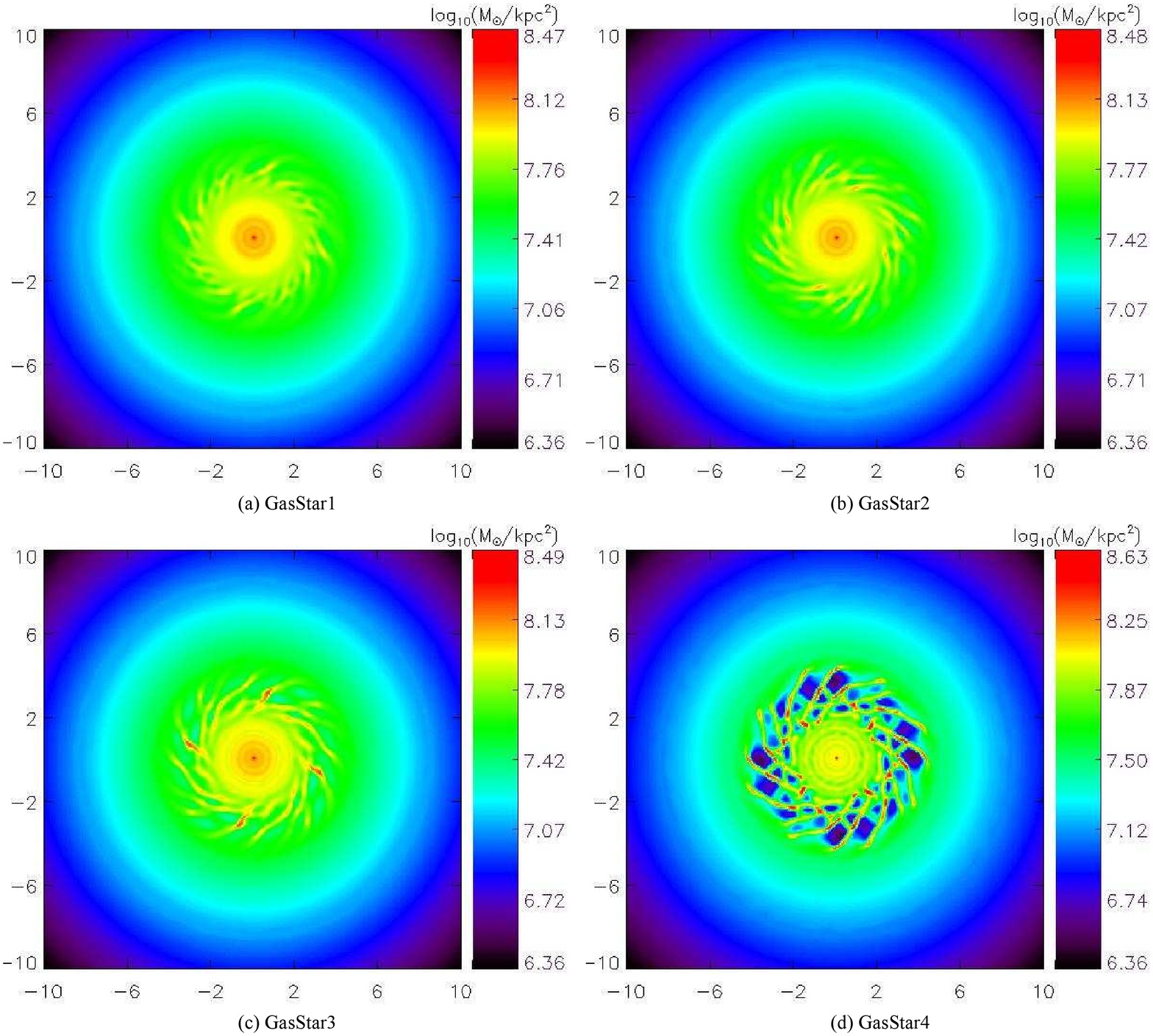,width=0.85\textwidth}}
}
\caption{Images (a) to (d) correspond to models GasStar1 to GasStar4.
They show the face-on surface density at $t=250$ Myr. The size of the 
images are 20 kpc $\times$ 20 kpc. The gas at the most unstable radius has
orbited around the center about two and half times. Spirals seen in
model GasStar2(b) and GasStar3(c) are due to swing amplification. In
(d) the disk fragments very fast mainly due to both the axisymmetric
instability and swing amplification. }
\label{fig:galprop}
\end{figure*}

The inclusion of a static stellar disk alters two important factors
which influence the stability of the disk. One is the rotation curve
and the other is the thickness of the gas disk. By changing the
rotation curve, the epicyclic frequency, $\kappa$, changes
accordingly. Supposing a flat rotation curve described by
$\Omega=V_0/R$, the epicyclic frequency $\kappa$ then reads:
\begin{equation}
\kappa^2=2\Omega^2=2\frac{V^2_0}{R^2},
\end{equation}
with $V_0$ being the rotation velocity. The presence of a stellar
disk tends to stabilize the gas disk via increasing $V_0$. However, by
increasing the gravitational pull in the vertical direction, the gas
disk becomes thinner and therefore more susceptible to gravitational
collapse. In Section 4.1, we have already seen that the scale-height,
which is governed by the temperature of the disk, is a very sensitive
factor for the disk stability. GasStar1 to GasStar4 are designed to
explore the competition between the two opponents.

From Fig. 6, we first notice that, compared to Fig. 4, the threshold
value, $Q_{\rm crit}$, is boosted from 0.693 to 0.75 due to the
decrease in scale-height. This makes the disk more prone to
gravitational instability. On the other hand, the change of the
rotation curve drastically shifts the dash-dotted curve
upwards. Instability only sets in once the temperature of the gas disk
drops below $T_{\rm crit} \sim 6000$ K. Overall, the presence of
the static stellar disk tends to stabilize the disks.

Figure 7 shows the surface density after an evolution of 250
Myr. During this period, the gas in the most unstable region has
finished 2.5 orbits. All the gas disks are developing multi-armed
spiral structures within the region where the disk is the most
vulnerable to instability according to Fig. 6. At this moment, the
most unstable disk, GasStar4, is experiencing fragmentation. High
density filaments are evident from the image. While GasStar2 is still
in its early stage of instability, GasStar3 is just about to enter the
fragmentation phase. GasStar1, on the other hand, does not fragments
at all during the course of simulation.

The trend is clear. The cooler the disk, the faster it
fragments. The spiral structure seen in these images are due to swing
amplification (Toomre 1981; Goldreich \& Lynden-Bell 1965), a
mechanism that is capable of amplifying the perturbation by swinging
the leading waves to trailing. Swing amplification is effective as the
disk $Q$ (dash-dotted line) is approaching the threshold $Q_{\rm th}$
(the solid line). The spirals are sheared, become tighter and tighter
and enhanced. Once the density reaches the supercritical point,
instability sets in.

\section{Spontaneously Induced Spiral Structure}
% fig. 8
\begin{figure*}
{
\centerline{\psfig{figure=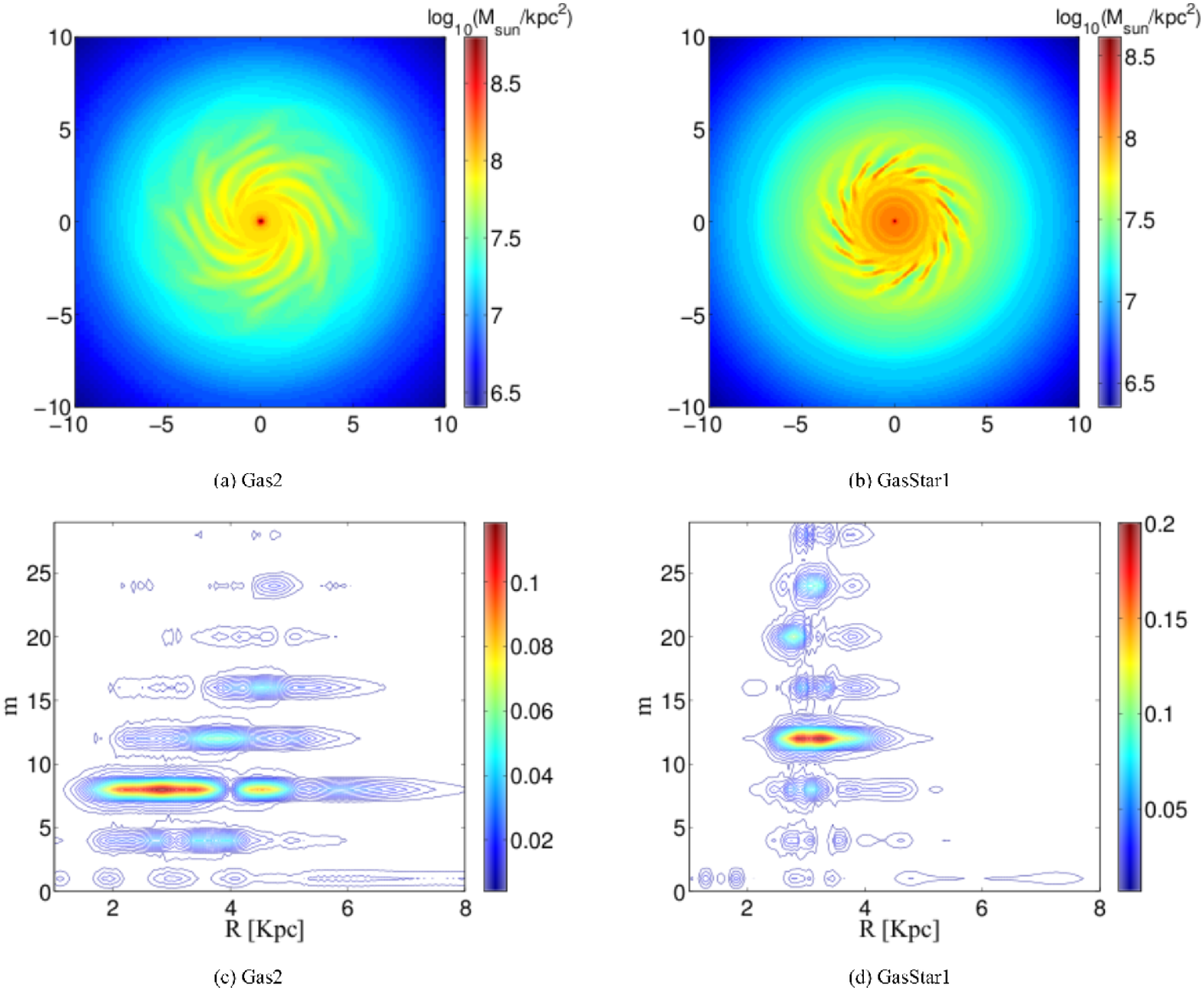,width=0.85\textwidth}}
}
\caption{The image size in (a) and (b) is 20 kpc $\times$ 20 kpc. (a)
The surface density of Gas2 at $t=750$ Myr. (b) The surface density of
GasStar1 at $t=500$ Myr. In both cases, the inner parts of the gas
disks, which have been evolved for about four orbital times,
developing spiral structure. Contour plots (c) and (d) are the Fourier
maps of (a) and (d), respectively.  In (c) and (d), the horizontal
axis represents radius, the vertical axis is the number of arms, $m$,
obtained by Fourier analysis. The color represents the intensity of
each Fourier mode, the redder the stronger. }
\label{fig:galprop}
\end{figure*}

An interesting feature which is hard to ignore in Fig. 5 and Fig. 7 is
that the marginally stable disks are spontaneously developing
multi-arm spiral structures. We have already seen in Section 3 and 4
that the effect of the disk thickness is to shift the range of the
marginally stable region downwards and therefore to stabilize the
disk. As we systematically lower the temperature to probe the onset of
instability, runs with as well as without stellar potential are experiencing
swing amplification.

Hohl (1971) found that disks which are marginally stable to
axisymmetric perturbation are prone to develop a large-scale bar
structure. This finding initiated both numerical (Zang \& Hohl 1978;
Sellwood 1981, 1985; Fuchs \& von Linden 1998; Sellwood \& Moore 1999)
and theoretical studies (Kalnajs 1978; Sawamura 1988; Vauterin \&
Dejonghe 1996; Pichon \& Cannon 1997; Evans \& Read 1998; Fuchs 2001)
of marginally stable disks. Goldreich \& Lynden-Bell (1965) and Toomre
(1981) pointed out that self-gravitating, differentially rotating
disks are able to amplify spiral waves by shearing a leading wave into
a trailing one. Three key ingredients, self-gravity, shearing and
epicyclic motions work harmonically to make the phenomenon now coined
with the name `swing amplification' happen.

Three necessary conditions need to be fulfilled in order to facilitate
the swing amplification (Toomre 1981; Fuchs 2001; Fuchs \& von Linden
1998; Binney \& Tremaine 2008). First, the disk must be marginally
stable, i.e., for an infinitesimally thin disk, $1<Q<2$, as defined by
Eq. (31). Second, the parameter $X=k_{\rm crit}R/m=k_{\rm crit}/k_y$
(Toomre 1981; Binney \& Tremaine 2008), with $m$ being the number of
arms and $k_{\rm crit}=\kappa^2/(2\pi G\Sigma_{\rm g})$ the critical
wave number, has to be of order unity, i.e., somewhere between 1 and 3
(Goldreich \& Lynden-Bell 1965; Julian \& Toomre 1966; Toomre
1981). Third, there must be a mechanism that is able to induce
leading arms in the system either explicitly by hand (Toomre 1981)
or implicitly by random fluctuation induced by numerical noise (Toomre
1990; Sellwood \& Carlberg 1984; Fuchs 2001). We notice that most of
these works mentioned above are for live stellar disks not directly
for the gas disk. But since the amplification principles are the same,
the results are still applicable to pure gas disks.

As shown in Fig. 8a and 8b, GasStar1 gets more arms than
Gas2 does. To be more quantitatively, Fig. 8c and 8d show the Fourier components
as a function of radius. They are obtained by doing Fourier transform to $(\Sigma_{\rm g}(R,\phi)-\overline{\Sigma_{\rm g}}(R))/ \overline{\Sigma_{\rm g}} (R)$, 
where $\overline{\Sigma_{\rm g}}(R)$ is the averaged surface density 
of a given radius. Note that the dominating modes tends to be multiples 
of $m=4$. This is a consequence of using a Cartesian grid, for which 
$m=4$ is the natural mode. However, the dominating mode is determined 
by physics. The dominating mode of Gas2 is $m=8$ while in GasStar1 $m=12$.

As is apparent from Eq. (38), including a stellar disk causes an
increase in $k_{\rm crit}$. Consequently, a larger value of $m$ is
required in order to satisfy $1 < X < 3$. From the image shown in
Fig. 8a and the relation, $k_{\rm crit}\propto \kappa^2$, to keep $X$
a constant, the number of spiral arms in GasStar1 can be crudely
estimated as $m\simeq 15$. More precisely, the number of spiral arms,
$m$, is predicted by (Toomre 1981; Athanassoula, Bosma \& Papaioannou
1987; Fuchs 2001, 2008):
\begin{equation}
m=\frac{2\pi R}{\lambda_{\rm{max}}}, 
\end{equation}
with $\lambda_{\rm{max}}$ being defined by:
\begin{equation}
\lambda_{\rm{max}}=\frac{\lambda_{\rm crit}}{\chi(A/\Omega)},
\end{equation}
where $\lambda_{\rm crit}=2\pi/k_{\rm crit}$. The coefficient $\chi$
is a function of rotation curve (Fuchs 2001), as measured by Oort's
constant $A$.

We employ Eq. (39) to analytically estimate the number of arms and
compare the predictions with the images shown in Fig. 8. For Gas2,
spirals appear between 2 kpc and 5 kpc. \revise{Within this radial
range, the most unstable wavelength ranges from 2.0 to 3.6 kpc.
The corresponding prediction for $m$ ranges from 6 to 9, while the 
simulation reveals a spiral pattern with 8-fold symmetry. For 
GasStar1, spirals are prominent between 3 and 4 kpc, while the 
corresponding most unstable wavelength ranges from 1.4 to 2.0 kpc. 
The twelve arms developing in GasStar1 should be compared to a 
predicted $m$ ranging from 13 to 14. Hence, overall the trends in the
simulations are in reasonable agreement with our predictions. Note
that the spatial resolution in both simulations ranges from 60 pc to
120 pc, indicating that the most unstable wavelengths are
well-resolved.}

The observed small deviations can be explained as follows. First, the
formulation used to predict the number of arms is precise only for
stellar disks. However, Toomre (1981) has shown the strikingly
similar behavior of gaseous disks (Goldreich \& Lynden-Bell 1965) and
stellar disks (Julian \& Toomre 1966). Therefore, we have confidence
that Eq. (39) is still applicable to gaseous disks. Second, the
number of arms has to be an integer, a number of fraction given by
Eq. (39) has no physical meaning. Third, the usage of a Cartesian
grid introduces the multiples of the natural $m=4$ mode, which
manifests itself in the Fourier transform of the surface
density. Fourth, swing amplification picks up the dominating
mode. It takes some time to fully develop the dominating mode. All
these factors combined determine the number of spiral seen in our
simulations. It is important to realize that the most unstable radius
according to the axisymmetric instability criterion might not be the
most effective site for swing amplification, since the shear plays an
important role in this process.
 
\revise{Without any external pumping source, spiral waves produced by
swing amplification should be a transient phenomenon. Similar to
material spirals, swing amplified spiral waves will experience
azimuthal shearing which reduces their pitch angle until they become
too tightly wound to be identified. As an example, in the Gas2
simulation, the spiral arm that appears around $R=2$ kpc initially has
a pitch angle of 90$^\circ$ and should be sheared to less than 1$^\circ$
within 2.2 Gyr. On the contrary, we find that the spontaneously
induced spirals seen in Gas2 can last for more than 3 Gyr and
still keep the pitch angle relatively open}. This result suggests at
least one mechanism keeps replenishing noise into the disk, leaving
the physics to pick up the dominating mode and sustain the waves. This
noise can be caused by numerics or preexisting waves.

\section{Summary}

In this paper we have developed a simple and effective method to
compute the three-dimensional density and velocity structure of an
isothermal gas disk in hydrodynamic equilibrium in the presence of an
arbitrary external potential (i.e., dark matter halo and/or stellar
disk). This is ideally suited to set-up the initial conditions of a
three-dimensional gas disk in equilibrium in hydrodynamical
simulations. We first notice that as long as the gas is barotropic or
has constant temperature at $t=0$, the circular velocity needed to
support the self-consistent disk is independent of the height above or
below the midplane. This feature greatly simplifies the process of
specifying the initial velocity field. All we need to know is the
rotation velocity in the midplane.

To specify the density distribution self-consistently, the hydrostatic
equation coupled with the reduced Poisson equation is adopted to
develop the vertical structure of the gas. Two sets of second-order
non-linear differential equations are found. One is directly
associated with the gas density called the density method, the other
associated with the gas potential called the potential method. In a
simulation involving a huge dynamic range (using AMR techniques), the
potential method is shown to be numerically more stable. A simple
local iteration can be performed to gain a better control on the shape
and the mass of disks. These ideas are simple enough to be
incorporated into any existing code, and most importantly they are
very effective.

With gas disks that are in detailed balance, we are able to
systematically investigate the axisymmetric stability of a fully
three-dimensional disk for the first time. We
probe the onset of instability both semi-analytically and
numerically. Simulations without stellar disk show that the thickness
of the gas disk, which is governed by the temperature of the disk, has
a huge impact on the disk stability. The reduction of the gravity
decreases the threshold value by around 30 percent in our models. As
we gradually lower the gas temperature, the threshold $Q_{\rm th}$
shifts up, the disk $Q$ shifts down, and the system starts to develop
multi-arm structure via swing amplification. The onset of the
instability in simulations matches the theoretical prediction very
well as shown in Fig. 4 and Fig. 5. The disk fragments as the two
curves, $Q$ and $Q_{\rm th}$, come very close to each other.

The influence of the stellar disk is less obvious. Its presence has a
stabilizing effect on the gas disk through changing the rotation curve
and a destabilizing one through the increase of the local
gravitational force. The simulation results show that overall the
presence of the stellar disk tends to stabilize the gas disk. But this
conclusion comes with a caveat. The interaction between live stars and
gas might be important. A live stellar disk itself can be unstable or
marginally stable. Perturbations from the interstellar medium can
trigger instabilities in the stellar disk. Since stars dominate the
mass budget in Milky-Way type galaxies (more than 90 percent), and
because gas is highly responsive and dissipative, the interplay
between both components is one of the most interesting subjects in
galactic dynamics. Tackling this problem needs elaborate initial
conditions for the live stellar disk or the combined disk. We stress
that the potential method developed in this paper is compatible with
the formulation in KD95. This makes the self-consistent combined disk
a natural direction for future work.

Marginally stable disks are susceptible to the process of swing
amplification, a prevalent mechanism that triggers self-induced
spirals. Simulations Gas2 and GasStar1 show the spirals are prominent
in the regions in which the gas can respond to swing
amplification. Semi-analytic result relates the most vulnerable
wavelength in azimuthal direction, $\lambda_\textrm{max}$, to the
number of arms. Numerically, The natural mode of a Cartesian grid
together with the swing amplification determine the dominating mode of
the spiral structure. Our numerical results with or without stellar
disk shows the correct characteristics of the swing amplification. It
happens in marginally stable disks and the number of arm fits
reasonably well to the analytic prediction. In the run of GasStar2,
swing amplification eventually leads to disk fragmentation once the
density becomes supercritical to the gravitational
instability. However, in a sub-critical case like Gas2, the spiral
structure can survive more than 3 Gyrs without fragmenting the disk,
suggesting at least one mechansim is sustaining the waves. The number
of arms suggests a characteristic wavelength relating to the upper
limit of the mass of giant molecular clouds (Escala 2008).
\label{sec: Summary}

\section*{Acknowledgments}

R.S.K.\ acknowledges financial support from the German {\em  
Bundesministerium f\"{u}r Bildung und Forschung} via the ASTRONET  
project STAR FORMAT (grant 05A09VHA) and from the {\em Deutsche  
Forschungsgemeinschaft} (DFG) under grants no.\ KL 1358/1, KL 1358/4,  
KL 1359/5, KL 1359/10, and KL 1359/11. R.S.K.\ furthermore thanks for  
subsidies from a Frontier grant of Heidelberg University sponsored by  
the German Excellence Initiative and for support from the {\em  
Landesstiftung Baden-W{\"u}rttemberg} via their program International  
Collaboration II (grant P-LS-SPII/18 ). R.S.K. also thanks the KIPAC at
Stanford University and the Department of Astronomy and Astrophysics 
at the University of California at Santa Cruz for their warm hospitality
during a sabbatical stay in spring 2010. H.-H.W.\ acknowledges the 
financial support from the International Max Planck Research School (IMPRS) 
Heidelberg and the technical support with RAMSES from Dr. R. Teyssier. 
Authors\ acknowledge the anonymous referee's comments and discussions 
that make this paper better. Numerical simulations were performed on 
the PIA cluster of the Max-Planck-Institut f\"ur Astronomie at the 
Rechenzentrum in Garching. 

%%%%%%%%%%%%%%%
% Bibliography
%%%%%%%%%%%%%%%

\appendix

\section[]{The Derivation of Rotation Velocity}

Equation (11) can be re-written as
\begin{eqnarray}
p(R,z)&=&\rho(R,z)\left.\frac{p(R)}{\rho(R)}\right|_{z=0}-\rho(R,z)\int^{z}_{0}\frac{p}{\rho^2}\frac{\partial \rho}{\partial z}{\rm d}z \nonumber \\
&-&\rho(R,z)[\Phi(R,z)-\Phi(R,z=0)], 
\end{eqnarray}
where we have replaced $\Phi_z=\Phi(R,z)-\Phi(R,z=0)$. Inserting
Eq. (A1) in Eq. (1) involves a partial derivative to the integral, let
us prepare it first:
\begin{eqnarray}
\lefteqn{\frac{\partial}{\partial R}\left(\int^{z}_{0}\frac{p}{\rho^2}\frac{\partial \rho}{\partial z}{\rm d}z\right) = \int^{z}_{0} \frac{\partial}{\partial R}\left( \frac{p}{\rho^2}\frac{\partial \rho}{\partial z}\right){\rm d}z} \nonumber \\
&=& \int^{z}_{0}\left\lbrace\frac{\partial^2}{\partial R \partial z}\left(\frac{p}{\rho^2}\rho \right)-\frac{\partial}{\partial R}\left[ \rho \frac{\partial}{\partial z}\left( \frac{p}{\rho^2}\right)\right] \right\rbrace {\rm d}z \nonumber \\
&=& \int^{z}_{0} \left\lbrace \frac{\partial}{\partial z}\left[ \frac{p}{\rho^2}\frac{\partial \rho}{\partial R}+\rho\frac{\partial}{\partial R}\left( \frac{p}{\rho^2}\right)\right]-\frac{\partial}{\partial R}\left[ \rho \frac{\partial}{\partial z}\left( \frac{p}{\rho^2}\right)\right]\right\rbrace {\rm d}z \nonumber \\
&=& \frac{p(R,z)}{\rho(R,z)}\frac{\partial \ln \rho(R,z)}{\partial R}-\frac{p(R,z=0)}{\rho(R,z=0)}\frac{\partial \ln \rho(R,z=0)}{\partial R} \nonumber \\
& & + \int^{z}_{0} \left\lbrace \left(\frac{\partial \rho}{\partial z}\right)\frac{\partial}{\partial R}\left( \frac{p}{\rho^2}\right)-\left(\frac{\partial \rho}{\partial R}\right)\frac{\partial }{\partial z}\left(\frac{p}{\rho^2} \right)\right\rbrace {\rm d}z
\end{eqnarray}
With Eq. (A2), the first term of Eq. (1) then becomes:

\begin{eqnarray}
\lefteqn{\frac{1}{\rho(R,z)}\frac{\partial p(R,z)}{\partial R}= \frac{\partial}{\partial R}\left(\frac{p(R,z=0)}{\rho(R,z=0)}\right)} \nonumber \\
&-& \left[ \frac{\partial \Phi(R,z)}{\partial R}-\frac{\partial \Phi(R,z=0)}{\partial R} \right] \nonumber \\
&+& \frac{p(R,z=0)}{\rho(R,z=0)}\frac{\partial \ln \rho(R,z=0)}{\partial R} \nonumber \\
&-& \int^{z}_{0} \left\lbrace \left(\frac{\partial \rho}{\partial z}\right)\frac{\partial}{\partial R}\left( \frac{p}{\rho^2}\right)-\left(\frac{\partial \rho}{\partial R}\right)\frac{\partial }{\partial z}\left(\frac{p}{\rho^2} \right)\right\rbrace {\rm d}z \nonumber \\
&+& \frac{\partial \ln \rho(R,z)}{\partial R}\left\lbrace \frac{p(R,z=0)}{\rho(R,z=0)} \right.
- \frac{p(R,z)}{\rho(R,z)} \nonumber \\ 
&-& \Phi(R,z) + \Phi(R,z=0)
- \left. \int^{z}_{0} \frac{p}{\rho^2}\frac{\partial \rho}{\partial z} {\rm d}z \right\rbrace
\end{eqnarray}
Equation (11) says that the term in the big brace should vanish. And
therefore, Eq. (1) reduces to
\begin{eqnarray}
\lefteqn{\frac{1}{\rho}\frac{\partial p}{\partial R}+\frac{\partial \Phi}{\partial R}=\frac{\partial}{\partial R}\left[\frac{p(R,z=0)}{\rho(R,z=0)}\right]} \nonumber \\
&+& \frac{\partial \Phi(R,z=0)}{\partial R}
+ \frac{p(R,z=0)}{\rho(R,z=0)}\frac{\partial \ln \rho(R,z=0)}{\partial R} \nonumber \\
&-& \int^{z}_{0} \left\lbrace \left(\frac{\partial \rho}{\partial z}\right)\frac{\partial}{\partial R}\left( \frac{p}{\rho^2}\right)-\left(\frac{\partial \rho}{\partial R}\right)\frac{\partial }{\partial z}\left(\frac{p}{\rho^2} \right)\right\rbrace {\rm d}z
\end{eqnarray}
For the barotropic gas, i.e., $p(\rho)$, the integrand of the integral
vanishes:
\begin{eqnarray}
\lefteqn{\left(\frac{\partial \rho}{\partial z}\right)\frac{\partial}{\partial R}\left( \frac{p}{\rho^2}\right)-\left(\frac{\partial \rho}{\partial R}\right)\frac{\partial }{\partial z}\left(\frac{p}{\rho^2} \right)} \nonumber \\
&=& \left( \frac{\partial \rho}{\partial z}\right)\frac{\partial}{\partial \rho}\left( \frac{p}{\rho^2}\right)\frac{\partial \rho}{\partial R}-\left( \frac{\partial \rho}{\partial R}\right)\frac{\partial}{\partial \rho}\left( \frac{p}{\rho^2}\right)\frac{\partial \rho}{\partial z}=0
\end{eqnarray}
For the cases of initially constant temperature, the specific internal
energy, $e$, is a constant and therefore the pressure is a function of
density only, the integrand vanishes.

\section[]{The effect of the disk thickness on the midplane potential}

For an axisymmetrically and infinitesimally thin disk, the potential
can be evaluated by the following relation (Binney \& Tremaine, 2008):
\begin{equation}
\Phi(R,z) = \int^{\infty}_{0}{\rm d}kS_0(k)J_0(kR)e^{-k|z|},
\end{equation}
where $J_0$ is the Bessel function of the first kind of order zero and
$S_0$ is the Hankel transform of $-2\pi G\Sigma_0$ defined by:
\begin{equation}
S_0(k) = -2\pi G \int^{\infty}_{0}{\rm d}R' R' J_0(kR')\Sigma_0(R')
\end{equation}
With Eq. (B1) and Eq. (B2), we can superimpose the potential
contributed by each gas layer. For the sake of simplicity, we assume
that the volume density has the double exponential profile:
\begin{equation}
\rho_0(R,z) = \Sigma_0 e^{-R/R_{\rm d}}\frac{e^{-z/h_z}}{2h_z}, 
\end{equation}
with $h_z$ being the scale-height of the gaseous disk, Eq. (B2) then
becomes (Gradshteyn \& Ryzhik 1965, hereafter GR65, 6.623-2):
\begin{eqnarray}
\lefteqn{S_0(k,z) = \frac{-2 \pi G \Sigma_0 e^{-z/h_z}}{2h_z}\Delta z \int^{\infty}_{0}{\rm d}R' R' J_0(kR')e^{-R'/R_d}} \nonumber \\
&=& \frac{-2 \pi G \Sigma_0 e^{-z/h_z}}{2h_z}\Delta z  \frac{\xi}{(\xi^2+k^2)^{3/2}},
\end{eqnarray}
with $\xi=1/R_{\rm d}$. $\Delta z$ represents the infinitesimal
thickness introduced to keep the dimension correct. The potential
which takes into account the thickness of the disk then reads:
\begin{eqnarray}
\Phi(R,z) &=& -2\pi G \Sigma_0\int^{\infty}_{0}{\rm d}k\frac{\xi}{(\xi^2+k^2)^{3/2}}J_0(kR)\nonumber \\
&\times& \int^{\infty}_{-\infty}e^{-k|z-h|}\frac{e^{-h/h_z}}{2h_z}{\rm d}h.
\end{eqnarray}
Evaluating the potential at the midplane, $z=0$, yields:
\begin{equation}
\Phi(R,z=0)=-2\pi G \Sigma_0\int^{\infty}_{0}{\rm d}k\frac{\xi}{(\xi^2+k^2)^{3/2}}J_0(kR)\frac{1}{1+kh_z}.
\end{equation}

Given the finite scale-height, the integral can be evaluated
numerically and compared with the result of the infinitesimally thin
disk.

\section[]{The Derivation of the reduction factor}

To derive the reduction factor $F$ defined by Eq. (35) we need to
evaluate the integral of the form:
\begin{eqnarray}
\lefteqn{\int^{\infty}_{-\infty}e^{-k|h|}\textrm{sech}^2(ah){\rm d}h=2\int^{\infty}_{0}e^{-kh}\textrm{sech}^2(ah){\rm d}h} \nonumber \\
&=& \frac{2}{a}\left\lbrace 1-\frac{k}{2a}\left[ H\left( \frac{k}{4a}\right)-H\left( \frac{k}{4a}-\frac{1}{2}\right)\right]\right\rbrace.
\end{eqnarray}
The last line can be reached by looking up the formulae 3.541, 8.370,
8.361-7 listed in the integral table (GR65) and the definition
Eq. (36). In the last line, we have employed the recursive relation
(8.365-1 GR65):
\begin{equation}
H(\alpha) = H(\alpha-1) + \frac{1}{\alpha}. 
\end{equation}
The asymptotic behavior of the harmonic number reads (8.367-2, 8.367-13 GR65):
\begin{equation}
 H(\alpha) = \ln\alpha + \gamma + \frac{1}{2}\alpha^{-1}-\frac{1}{12}\alpha^{-2}+\frac{1}{120}\alpha^{-4}+ O(\alpha^{-6}),
\end{equation}
with $\gamma=0.5772156649$ (8.367-1 GR65) being the Euler-Mascheroni
constant. Note that Eq. (C3) is only reliable when $\alpha \geq 1$. We
employ the recursive relation (C2) to evaluate $H(\alpha)$ for
$-1<\alpha<1$.

\section[]{The vertical force ratio}

\begin{figure}
\begin{minipage}[c]{0.5\textwidth}
\psfig{figure=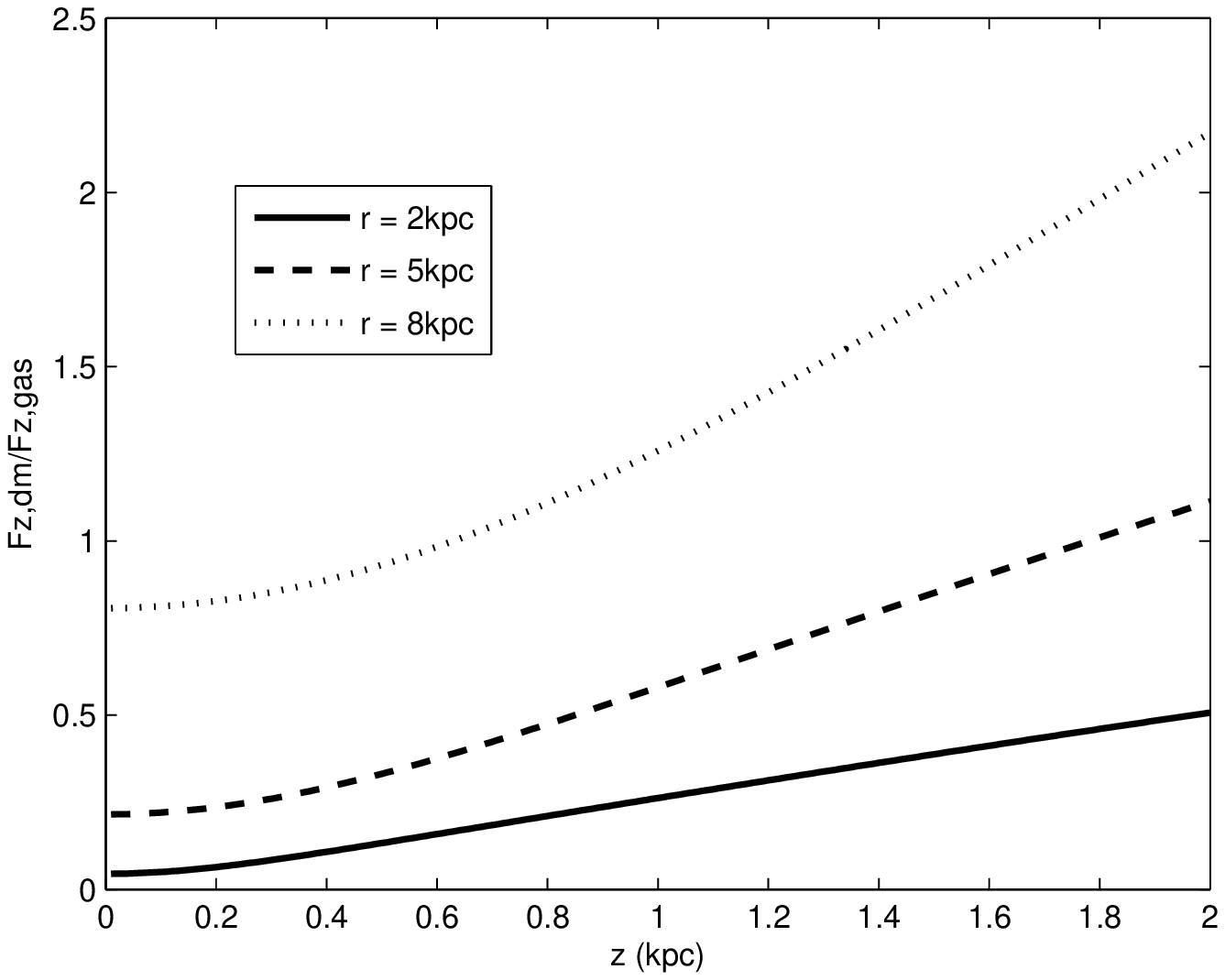,width=\textwidth}
\end{minipage}
\caption{The force ratio $F_{z,{\rm DM}}/F_{z,{\rm gas}}$ at R=2, 5
  and 8 kpc. It shows that the vertical structure of the inner disk is
  determined mostly by the self-gravity of gas.}
\end{figure}

The vertical force ratio measures the impact of the halo force on the
vertical structure. The simplified Poisson equation for isothermally
self-gravitating gas disk reads:
\begin{equation}
\frac{\partial^2 \Phi_{\rm g}}{\partial z^2}=4\pi G \rho_0(R) \textrm{sech}^2\left(\frac{z}{h_z}\right),
\end{equation}
where $h_z$ being a measure of the scale-height. Parameter $h_z$ can
be related to the volume density in the midplane, $\rho_0(R)$ by:
\begin{equation}
h_z=\sqrt{\frac{c^2_{\rm s}}{2\pi G\rho_0}}.
\end{equation}
The corresponding vertical force for the gas then becomes:
\begin{equation}
F_{z,{\rm gas}} = -\frac{\partial \Phi}{\partial z}=-4\pi G h_z \rho_0(R){\tanh(z/h_z)}.
\end{equation}
For a NFW halo, the vertical force can be written down directly:
\begin{equation}
F_{z,{\rm DM}} = \frac{GM_{200}}{f(c)}\left(\frac{c}{r_{200}} \right)^2\frac{x/(1+x)-\ln(1+x)}{x^2}\frac{z}{\sqrt{R^2+z^2}},
\end{equation}
with $x=c\sqrt{R^2+z^2}/r_{200}$. Figure D1 then shows the force ratio
$F_{z,{\rm DM}}/F_{z,{\rm gas}}$ as a function of vertical height
$|z|$ at different radii. Comparing to the rotation curve shown in the
left panel of Fig. 1, although the dynamics is still dictated by the
potential of the dark halo, the vertical structure of the gaseous disk
is mainly determined by the self-gravity of the gas
component. However, it raises another issue, the presence of the
stellar disk will dominate both the dynamics and the vertical
structure of the gas and will affect the stability of the gas
component via changing the thickness of the gaseous disk and the
rotation curve.

\section[]{Validity check of the reduced Poisson Equation for the gas disk}

\begin{figure}
\begin{minipage}[c]{0.5\textwidth}
\psfig{figure=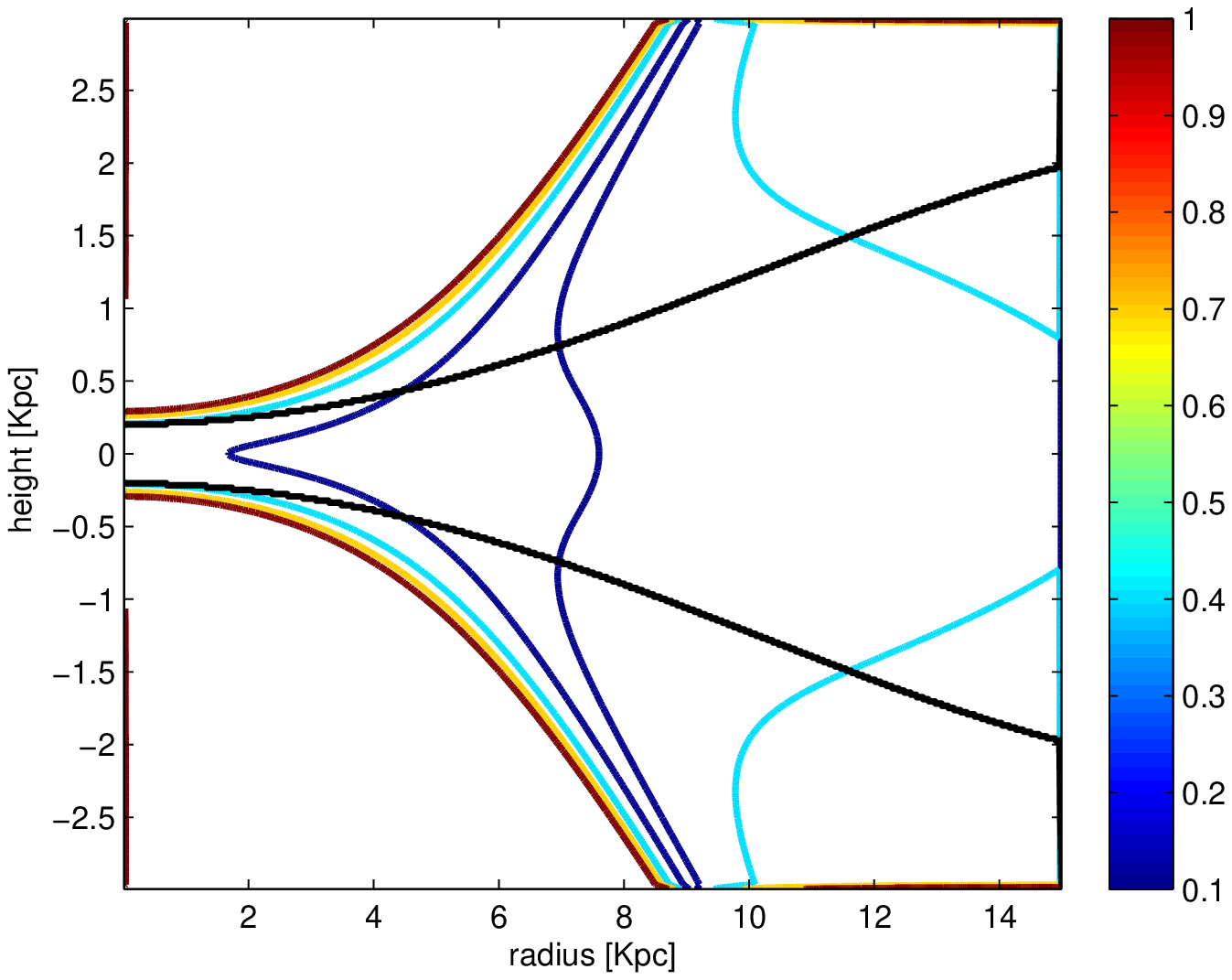,width=\textwidth}
\end{minipage}
\caption{Contour map of $\epsilon$. The black lines
  represent the scale-height of the gas disk. }
\end{figure}

\revise{Throughout this paper we have assumed that the radial potential
gradients of the disk are negligible compared to the vertical
gradients, such that the Poisson equation reduces to Eq.~(14). We now
test this assumption by computing the ratio}
\begin{equation}
\epsilon \equiv |\frac{1}{R}\frac{\partial}{\partial R}
(R\frac{\partial \Phi_{\rm g}}{\partial R}) / 
\frac{\partial^2 \Phi_{\rm g}}{\partial z^2}|,
\end{equation}
\revise{with $\Phi_{\rm g}$ the gravitational potential of the gas disk. For a
realistic, analytical disk model, our assumption will be valid
as long as $\epsilon \ll 1$.}

\revise{Consider the Miyamoto \& Nagai (1975) potential:}
\begin{equation}
\Phi_{\rm g}(R,z) = -\frac{GM_{\rm g}}{\sqrt{R^2+(a+\sqrt{z^2+b^2(R)})^2}}\,.
\end{equation}
\revise{Here $a$ is a constant that controls the scale-length of the disk and
$b(R)$, which we take to be a function of radius, modulates the
scale-height of the disk. In the limit $b \rightarrow 0$ this model
reduces to the infinitesimal Kuzmin disk (e.g., Binney \& Tremaine
2008).  In an attempt to model the gas disk in our simulation `Gas0',
we adopt $a = 3.5$ kpc. In order to mimic the flaring of the Gas0
disk (see Fig.~2b), we consider}
\begin{equation}
b(R) = -1.58\times 10^{-5} R^4 + 1.21\times 10^{-2} R^2 + 0.20.
\end{equation}
\revise{Using the Poisson equation to solve (numerically) for the
corresponding density distribution yields the radial-dependent
scale-height shown as the solid black lines in Fig.~E1, and which is
comparable to that of the Gas0 disk.  The contours in Fig.~E1
are defined by constant values of $\epsilon$. These show that our
assumption that $\epsilon \ll 1$ is well-justified in the inner part
of the disk, out to $\sim 3$ scale-lengths, which encloses most of the
disk mass. The assumption that $\epsilon \ll 1$ deteriorates at larger
radii and at higher altitude away from the midplane. This might be in
part responsible for the very slight outward drifting of the disk seen in
Fig.~2b. In cases that include a stellar potential and/or cooler gas,
the gas disk is even thinner than the case considered here, resulting
in values for $\epsilon$ that are even smaller. Based on these
results, and based on the absence of significant disk thickening in
our simulations, we are confident that Eq.~(14) is sufficiently
accurate for all realistic gas disks.}


\begin{thebibliography}{}

\bibitem[]{Age09} 
Agertz O., Lake G., Teyssier R., Moore B., Mayer L., Romeo A.B., 2009, \mnras, 392, 294

\bibitem[]{Ath87}
Athanassoula E., Bosma A., Papaioannou S. 1987, A\&A, 179, 23

\bibitem[]{Bin87} 
Binney J., Tremaine S. 2008, Galactic Dynamics (Princeton: Princeton Univ. Press, 2nd Ed.)

\bibitem[]{Cas83}
Casertano S., 1983, \mnras, 203, 735

\bibitem[]{Esc08}
Escala A., Larson R.B., 2008, \apj, 685, L31

\bibitem[]{Eva98}
Evans N.W., Read J.C.A., 1998, \mnras, 300, 106

\bibitem[]{Fuc01}
Fuchs B., 2001, A\&A, 368, 107

\bibitem[]{Fuc08}
Fuchs B., 2008, Invited contribution to Galactic and Stellar Dynamics in the era of high resolution surveys, Strassburg, March 16-20 [arXiv:0810.3503]

\bibitem[]{Fuc98}
Fuchs B., von Linden S., 1998, \mnras, 294, 513

\bibitem[]{Gam01}
Gammie C.F., 2001, \apj, 553, 174

\bibitem[]{Gol65b}
Goldreich P., Lynden-Bell D., 1965, \mnras, 130, 125

\bibitem[]{Gra65}
Gradshteyn I. S., Ryzhik I. M., 1965, Table of Integrals, Series, and Products, 4th Ed. (Academic Press, New York)

\bibitem[]{Hol71}
Hohl F. 1971, \apj, 168, 343

\bibitem[]{Jac74}
Jackson, P.D., Kellman, S.A., 1974, \apj, 190, 53

\bibitem[]{Jul66}
Julian W.H., Toomre A., 1966, \apj, 146, 810 (JT)

\bibitem[]{Kal78}
Kalnajs A.J., 1978, In IAU Symposium 77, Structure and Properties of Nearby Galaxies, ed. Berkhuijsen E.M.,   Wielebinski R.  (Dordrecht: Reidel), 133

\bibitem[]{Ken91}
Kent, S. M., Dame, T. M., Fazio, G. 1991, \apj, 378, 131

\bibitem[]{Kim07}
Kim W.T., Ostriker E.C., 2007, \apj, 660, 1232

\bibitem[]{Kim06}
Kim W.T., Ostriker E.C., 2006, \apj, 646, 213

\bibitem[]{Kim02}
Kim W.T., Ostriker E.C., Stone J.M., 2002a, \apj, 581, 1080

\bibitem[]{Kim02}
Kim W.T., Ostriker E.C., 2002b, \apj, 570, 132

\bibitem[]{Kui95}
Kuijken K., Dubinski J., 1995, \mnras, 277, 1341

\bibitem[]{Ler08}
Leroy A.K., Walter F., Brink E., Bigiel F., de Blok W.J.G., Modore B., Thornley M.D. 2008, AJ, 136, 2782

\bibitem[]{Li05a}
Li X., Mac Low M.-M, Klessen R.S., 2005a, \apj, 620, 19

\bibitem[]{Li05b}
Li X., Mac Low M.-M, Klessen R.S., 2005b, \apj, 626, 823

\bibitem[]{Li06}
Li X., Mac Low M.-M, Klessen R.S., 2006, \apj, 639, 879

\bibitem[]{Lin64}
Lin C.C, Shu F.H., 1964, \apj, 140, 646

\bibitem[]{Lis09}
Lisker T., Fuchs B., 2009, A\&A, 501, 429

\bibitem[]{Loc84}
Lockman, F. J. 1984, \apj, 283, 90

\bibitem[]{Miy75}
Miyamoto M., Nagai R., 1975, Astron. Soc. Japan, 27, 533

\bibitem[]{Nar02}
Narayan C.A., Jog C.J., 2002, A\&A., 394, 89

\bibitem[]{Nav97}
Navarro J., Frenk C., White S.D.M., 1997, \apj, 490, 493

\bibitem[]{Pic97}
Pichon C., Cannon R.C., 1997, \mnras, 291, 616

\bibitem[]{Raf01}
Rafikov R. R., 2001, \mnras, 323, 445

\bibitem[]{San84}
Sanders, D. B., Solomon, P. M., Scoville, N. Z. 1984, \apj, 276, 182

\bibitem[]{Saw98}
Sawamura M., 1988, PASJ, 40, 279

\bibitem[]{Sel99}
Sellwood J.A., Moore E.M., 1999, \apj, 510, 125

\bibitem[]{Sel85}
Sellwood J.A., 1985, \mnras, 217, 127

\bibitem[]{Sel84}
Sellwood J.A., Carlberg R.G., 1984, \apj, 282, 61

\bibitem[]{Sel81}
Sellwood J.A., 1981, A\&A, 99, 362

\bibitem[]{She06}
Shetty R., Ostriker E.C., 2006, \apj, 647, 997

\bibitem[]{Shu69}
Shu F.H., 1969, 158, 505

\bibitem[]{Spi42}
Spizter L., JR., 1942, \apj, 95, 329

\bibitem[]{Spr05}
Springel V., Matteo T.D., Hernquist L., 2005, \mnras, 361, 776

\bibitem[]{Tas06}
Tasker E., Bryan G., 2006, \apj, 641, 878

\bibitem[]{Tey02}
Teyssier R., 2002, A\&A, 385, 337

\bibitem[]{Too90}
Toomre A., 1990, in Dynamics and Interactions of Galaxies, ed. R. Wielen (Springer, Berlin), 292

\bibitem[]{Too81}
Toomre A., 1981, in Structure and Evolution of Normal Galaxies, ed. Fall, S.M., Lynden-Bell, D. (Cambridge: CUP), 111

\bibitem[]{Too64}
Toomre A., 1964, \apj, 139, 1217

\bibitem[]{van81a}
van der Kruit, P. C., Searle, L. 1981a, A\&A, 95, 105

\bibitem[]{van81b}
van der Kruit, P. C., Searle, L. 1981b, A\&A, 95, 116

\bibitem[]{Vau96}
Vauterin P., Dejonghe H., 1996, A\&A, 313, 465

\bibitem[]{Wou90}
Wouterloot, J. G. A., Brand, J., Burton, W. B., Kwee, K. K. 1990, A\&A, 230, 21

\bibitem[]{Zan78}
Zang T.A., Hohl F. 1978, \apj, 226, 521

\end{thebibliography}
\end{document}